\documentclass[%
 reprint,
 amsmath,amssymb,
 aps,pra,reprint,superscriptaddress,amsmath,amssymb, floatfix
]{revtex4-2}

\usepackage{graphicx}
\usepackage{dcolumn}
\usepackage{bm}
\usepackage{physics}
\usepackage{amsmath,amssymb}
\usepackage{amsfonts,fixmath}
\usepackage{xcolor}
\usepackage{comment}
\usepackage{mathtools}
\usepackage{pict2e}

\makeatletter

\newcommand{\bigcomp}{%
  \DOTSB
  \mathop{\vphantom{\sum}\mathpalette\bigcomp@\relax}%
  \slimits@
}
\newcommand{\ketbraind}[3]{\ket{#1}_#2\! \bra{#3}}

\newcommand{\bigcomp@}[2]{%
  \begingroup\m@th
  \sbox\z@{$#1\sum$}%
  \setlength{\unitlength}{0.9\dimexpr\ht\z@+\dp\z@}%
  \vcenter{\hbox{%
    \begin{picture}(1,1)
    \bigcomp@linethickness{#1}
    \put(0.5,0.5){\circle{1}}
    \end{picture}%
  }}%
  \endgroup
}
\newcommand{\bigcomp@linethickness}[1]{%
  \linethickness{%
      \ifx#1\displaystyle 2\fontdimen8\textfont\else
      \ifx#1\textstyle 1.65\fontdimen8\textfont\else
      \ifx#1\scriptstyle 1.65\fontdimen8\scriptfont\else
      1.65\fontdimen8\scriptscriptfont\fi\fi\fi 3
  }%
  \newcommand{\mycomment}[1]{}
}
\newtheorem{definition}{Definition}[section]
\newtheorem{lemma}{Lemma}[section]

\begin{document}
\preprint{APS/123-QED}

\title{Variational quantum thermalizers based on weakly-symmetric nonunitary multi-qubit operations}

\author{Elias Zapusek}
\email{zapuseke@ethz.ch}
\affiliation{Institute for Quantum Electronics, ETH Z\"urich, 8093 Z\"urich, Switzerland}

\author{Kristina Kirova}
\affiliation{Institute for Quantum Electronics, ETH Z\"urich, 8093 Z\"urich, Switzerland}
\affiliation{Institute for Integrated Circuits and Quantum Computing,
Johannes Kepler University Linz, 4040 Linz, Austria
}

\author{Walter Hahn}
\affiliation{Fraunhofer  Institute for Applied Solid State Physics IAF, Tullastr. 72, 79108 Freiburg, Germany}

\author{Michael Marthaler}
\affiliation{HQS Quantum Simulations GmbH, Rintheimer Str. 23, 76131 Karlsruhe, Germany}

\author{Florentin Reiter}%
\affiliation{Institute for Quantum Electronics, ETH Z\"urich, 8093 Z\"urich, Switzerland}
\affiliation{Fraunhofer  Institute for Applied Solid State Physics IAF, Tullastr. 72, 
79108 Freiburg, Germany}

\date{\today}

\begin{abstract}
We propose incorporating multi-qubit nonunitary operations in Variational Quantum Thermalizers (VQTs). VQTs are hybrid quantum-classical algorithms that generate the thermal (Gibbs) state of a given Hamiltonian, with applications in quantum algorithms and simulations. However, current algorithms struggle at intermediate temperatures, where the target state is nonpure but exhibits entanglement. We devise multi-qubit nonunitary operations that harness weak symmetries and thereby improve the performance of the algorithm.
Utilizing dissipation engineering, we create these nonunitary multi-qubit operations without the need for measurements or additional qubits. To train the ansatz, we develop and benchmark novel methods for entropy estimation of quantum states, expanding the toolbox for quantum state characterization. We demonstrate that our approach can prepare thermal states of paradigmatic spin models at all temperatures. 
Our work thus creates new opportunities for simulating open quantum many-body systems.
\end{abstract}

\maketitle

\section{Introduction}
The simulation of quantum-mechanical systems is one of the most promising applications of quantum computers \cite{feynman_Simulating_1982}. A common question is finding the ground state of a physical system \cite{peruzzo_variational_2014}, yet at present it is still debated whether quantum computers can calculate ground state properties faster than classical computers \cite{begusic_Fast_2024}. The calculation of equilibrium properties, which naturally arise in physical systems, is a promising avenue \cite{chen_efficient_2023}. In practice, any quantum system is never truly isolated but interacts with its environment. This interaction typically leads to the formation of Gibbs states, which are mixed states that describe a quantum system in thermodynamic equilibrium with its surroundings. These states are central to quantum statistical mechanics and have applications in quantum chemistry \cite{cao_quantum_2019,mazzola_phase_2018}, and the simulation of materials \cite{gull_superconductivity_2013,bauer_quantum_2020}. Gibbs states serve as resource states in quantum algorithms such as semidefinite programming \cite{brandao_quantum_2017,watts_quantum_2023}, and quantum machine learning \cite{verdon_quantum_2019,zoufal_variational_2021,amin_Quantum_2018}. Furthermore, they have interesting connections to quantum complexity theory \cite{bravyi_quantum_2022}.

The pervasiveness of thermal states drives the development of classical, near-term quantum \cite{martyn_product_2019,verdon_quantum_2019,chowdhury_Variational_2020,zoufal_variational_2021,wang_variational_2021,warren_Adaptive_2022,foldager_noise-assisted_2022, consiglio_variational_2025,ilin_dissipative_2025,fromm_simulating_2023,wang_symmetry_2023,selisko_extending_2024,polla_Quantum_2021,holmes_quantum_2022} and fault-tolerant quantum methods \cite{temme_quantum_2011,yung_quantumquantum_2012,wocjan_szegedy_2023,rall_thermal_2023,chen_efficient_2023,chen_quantum_2023,rouze_efficient_2024} to prepare and understand these states. 
Certain algorithms can prepare the thermal state of the model as the steady state and are provably efficient at high temperatures \cite{chen_efficient_2023,rouze_efficient_2024}. However, the use of resource-intensive subroutines and ancilla qubits renders the algorithms impractical for current and near-term quantum computers.
To address this challenge, new algorithms tailored for near-term devices are being developed. A key example are Variational Quantum Thermalizers (VQTs); these hybrid quantum-classical algorithms leverage parameterized quantum circuits to prepare thermal states \cite{martyn_product_2019,verdon_quantum_2019, chowdhury_Variational_2020, wang_variational_2021, zoufal_variational_2021, warren_Adaptive_2022, foldager_noise-assisted_2022,wang_symmetry_2023, fromm_simulating_2023,selisko_extending_2024,consiglio_variational_2025,ilin_dissipative_2025}. They usually focus more on practical resource demands suited to current quantum hardware rather than on provable theoretic guarantees.

Dissipation is a plentiful resource on present-day quantum computers and plays a key role in thermalization \cite{preskill_quantum_2018, scarani_thermalizing_2002}. Present-day algorithms underutilize its potential for the preparation of thermal states \cite{foldager_noise-assisted_2022,ilin_dissipative_2025}.
On the contrary, dissipation often hampers their performance, as many prepare a purification of the thermal state, which is a large entangled state and whose preparation is strongly affected by dissipation \cite{wu_variational_2019,verdon_quantum_2019, chowdhury_Variational_2020, wang_variational_2021, zoufal_variational_2021, holmes_quantum_2022, warren_Adaptive_2022, consiglio_variational_2025}. 

Here we present an algorithm which prepares the thermal states directly using the dissipation present in the system. This eliminates the need for additional ancilla qubits. Our approach builds on dissipative methods for ground state preparation \cite{kraus_Preparation_2008,reiter_scalable_2016}, variational \cite{foldager_noise-assisted_2022} and fault-tolerant methods for thermal state preparation \cite{chen_efficient_2023}. For the first time in parameterized quantum algorithms, we use weak and strong symmetries to design problem-specific unitary and nonunitary circuit elements, thereby simplifying the optimization problem. The natural cost function for the preparation of thermal states, the free energy, cannot be efficiently evaluated for arbitrary states. To create a cost function that can be evaluated efficiently we develop novel methods for estimating the entropy of quantum states and rigorously analyze their effectiveness growing the toolbox of quantum simulation techniques. Our advancements allow for the preparation of thermal states even for frustrated spin models at all temperatures.

\begin{figure*}[t!]
    \centering
    \includegraphics[width=\textwidth]{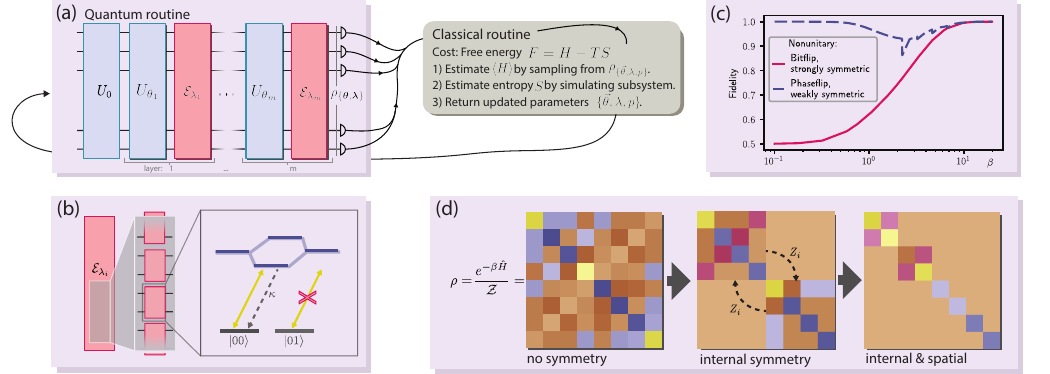}
    \caption{Overview.
    (a) 
    The hybrid algorithm. The quantum circuit begins with an initial unitary layer $U_0$, followed by $m$ layers, which are composed of unitary operations $U_{\theta_j}$ interleaved with nonunitary layers $\mathcal{E}_{\lambda_j}$. The parameters of all layers are trained using the free energy as the cost function. The energy term is measured by sampling from the output of the circuit, while the entropy is estimated using subsystems.
    (b) Nonunitary circuit elements. The nonunitary layers $\mathcal{E}_{\lambda_j}$ can be constructed from a product of single-qubit channels or more complex multi-qubit nonunitary operations. The multi-qubit nonunitary gates are realized via dissipation engineering. Typically, this involves a weak drive to an excited state manifold that experiences differential shifts due to a strong coupling within the excited states.
    The operation is made nonunitary by the decay of an auxiliary degree of freedom.
    (c) Weak and strongly symmetric channels. Breaking the strong symmetry is essential when one prepares Gibbs states at nonzero temperatures. Here we prepare thermal states of the transverse-field Ising model on a ring with internal symmetry, $R = \prod X_i$.
    The strongly symmetric channel can only reach fidelity $0.5$ for high temperatures while a weakly symmetric channel does not have this issue. Considering the symmetry sectors allows us to better understand the preparation.
    (d) Symmetry sectors. Visualization of the Gibbs state of the TFIM for three qubits. Using symmetries the problem can be simplified. The internal symmetry $R$ splits the space that needs to be explored into two symmetry sectors. The Gibbs state is a mixture of both symmetry sectors but there are no coherences between them. Applying a $Z$ operation on a single spin swaps the sectors. In panel (c) the ansatz utilizing the bitflip channel would be constrained to one of these symmetry sectors and therefore reaches a fidelity of 0.5 for high temperatures. Considering the spatial symmetries (here cyclic symmetry) further simplifies the problem. 
    }
    \label{fig:Story_double}
\end{figure*}

In Sec. \ref{sec:ansatz} we introduce the structure of the ansatz, which incorporates both unitary and nonunitary operations. The implementation of these nonunitary operations is detailed in Sec. \ref{sec:nonunitaryoperations}. Section \ref{sec:cost} presents the cost function and a method for estimating the entropy of quantum states essential for the evaluation of the cost function. The crucial tool to design the nonunitary operations, namely weak and strong symmetries, is discussed in Sec. \ref{sec:symsec}. These building blocks are combined in Sec. \ref{sec:spinthermal} to prepare thermal states of multiple spin models. Finally, we summarize our findings and offer perspectives for future research in Sec. \ref{sec:Conclusion}.

\section{Variational quantum thermalizers}\label{sec:ansatz}
Variational quantum thermalizers are hybrid quantum-classical algorithms that aim to prepare the Gibbs state of a system described by Hamiltonian $H$. The Gibbs state is given by 
\begin{equation}
	\rho_G = \frac{e^{-\beta H}} {\mathcal{Z}},
\end{equation} 
where $\beta = 1/(k_B T)$ is the inverse temperature and $\mathcal{Z} = \mathrm{Tr}[e^{-\beta H}]$ is the partition function. It is a probabilistic mixture of energy eigenstates of the Hamiltonian.

We generate the target mixed state by combining parameterized unitary quantum gates, with parameterized nonunitary channels. 
The unitary and nonunitary layers are alternated $m$ times, beginning with an initial unitary layer $U_0$. The resulting circuit is depicted in Fig. \ref{fig:Story_double} (a). For the sake of clarity, we incorporate these channels into the quantum circuit diagram.

The unitary layers are inspired by the quantum approximate optimization (QAOA) algorithm \cite{farhi_quantum_2014}.
The unitary part of the ansatz is composed of alternating layers of unitary evolution governed by the mixing Hamiltonian $H_M$, and the problem Hamiltonian $H$
\begin{align}\label{eq:UnitaryLayer}
    U_{\bm{\theta_j}} = e^{ -i\theta_{j,0} H_M } e^{ -i\theta_{j,1} H },
\end{align}
where $\bm{\theta}_j$, represents the set of parameters for the $j$-th layer. The initial state is chosen as the ground state of the mixing Hamiltonian, prepared by applying initial unitary $U_0$. This unitary ansatz can be viewed as a Trotterization of adiabatic state preparation. 

To prepare the desired mixed states, the algorithm incorporates nonunitary circuit elements, represented by quantum channels, which are alternated with unitary layers. We can express the ansatz for the algorithm as:
\begin{equation}\label{eq:Ansatz}
    \mathcal{A}_{\bm{\theta} , \bm{\lambda}} = \bigcomp_{j=1}^m \mathcal{E}_{\lambda_j} \circ \mathcal{U}_{\theta_j} .
\end{equation}
Here, $ \bigcomp$ represents the concatenation of the channels, $ \mathcal{U}_{\theta_j}( \rho) = U_{\theta_j}^\dag \rho \;U_{\theta_j} $ denotes the unitary operations parameterized by $ \theta_j $, and $ \mathcal{E}_{\lambda_j} $ represents the nonunitary quantum channels parameterized by $\lambda_j$.

 The circuit outputs the state $\rho_{\{\bm{\theta} , \bm{\lambda}\} } = \mathcal{A}_{\bm{\theta} , \bm{\lambda}} (U_0 \ketbra{0}{0}^{\otimes n} U_0^\dag ) $. By iteratively updating the circuit and channel parameters, we aim to prepare the Gibbs state.

\section{Physical implementation of nonunitary elements}\label{sec:nonunitaryoperations}
Our ansatz contains quantum channels; operations that go beyond the native gate set of conventional quantum computers. By the Choi isomorphism, any such channel can be decomposed into a unitary that involves additional degrees of freedom \cite{jamiolkowski_linear_1972,choi_completely_1975}. Both a large number of unitary operations and ancilla qubits are scarce resources on present-day quantum computers, making the Choi isomorphism impractical to implement.

Previous works consider channels that can be realized by probabilistic applications of unitaries \cite{verdon_quantum_2019,foldager_noise-assisted_2022}. Such channels can be readily implemented on any quantum computer, without a qubit overhead. However, they are unital, making the fully mixed state their steady state.

Non-unital channels differ fundamentally from unital channels. Non-unital channels can increase the purity of quantum states and can have more interesting fixed points such as entangled states \cite{reiter_scalable_2016, kastoryano_dissipative_2011, lin_dissipative_2013, morigi_dissipative_2015, cole_dissipative_2021, cole_resource-efficient_2022, malinowski_generation_2021, doucet_high_2020,malinowski_generation_2021}.
In the context of parameterized quantum circuits unital channels lead to noise-induced barren plateaus while non-unital channels can help avoid them \cite{singkanipa_beyond_2024}.

Dissipation is fundamentally nonunitary, which makes it ideal for generating irreversible operations. The discipline of dissipation engineering, also known as reservoir engineering, leverages the interaction between a quantum system and environmental degrees of freedom to perform quantum information processing tasks. \cite{harrington_engineered_2022, kronwald_dissipative_2014, agarwal_strong_2016, reiter_scalable_2016, kastoryano_dissipative_2011, lin_dissipative_2013, morigi_dissipative_2015, cole_dissipative_2021, cole_resource-efficient_2022, malinowski_generation_2021, rojkov_Bias_2022,doucet_high_2020, barreiro_open-system_2011, reiter_dissipative_2017, wang_autonomous_2019,rojkov_stabilization_2024}. Dissipation towards the environment lifts the requirement for classical measurement and feedback and has scaling and robustness advantages over unitary approaches \cite{kastoryano_dissipative_2011, lin_dissipative_2013, morigi_dissipative_2015}.

We sketch the realization of such a channel using dissipation engineering in Fig. \ref{fig:Story_double} (b) The layer $\mathcal{E}_{\lambda_i}$ is generated by multiple quasi-local jumps. One jump consists of excitation to an excited manifold which is tuned in and out of resonance depending on the multiqubit state. This enables us to excite only selected states. Subsequently, the excited state decays, completing the jump. Such operations can be readily implemented in physical systems such as trapped ions \cite{van_mourik_experimental_2024}, superconducting qubits \cite{reiter_steady-state_2013}, or quantum dots in cavities \cite{zapusek_nonunitary_2023}. Furthermore, the operational primitives have been shown to be universal for implementing Lindbladians on two qubits \cite{van_mourik_experimental_2024}. The full details are given in \ref{ap:jump_experiment}.

\section{Cost function and entropy approximation}\label{sec:cost}
The circuit parameters are optimized to prepare a desired thermal state by minimizing a cost function that captures how well the circuit prepares the target state. For the preparation of thermal states, the free energy is a natural cost function as Gibbs states minimize the free energy of the system,
\begin{equation}
	F(\rho) = \beta  \langle H \rangle_{\rho} - S(\rho).
\end{equation}
where the first term is the expectation value of the Hamiltonian multiplied by the inverse temperature, $\beta$. It captures the energy of the state, $E = \langle H \rangle = \mathrm{Tr}[H\rho]$, which for models consisting only of local Pauli strings can be efficiently computed on a quantum computer \cite{peruzzo_variational_2014}. The second term is the von Neumann entropy, $S = - \mathrm{Tr}[ \rho \log \rho]$. Evaluating it presents a challenge, as the entropy is not a linear function (observable) of the quantum state and, therefore, would require tomography to be evaluated exactly. 
This is infeasible for moderately sized systems as the sample complexity of tomography, and consequently of entropy estimation, scales exponentially with system size. To address this, we explore and benchmark alternative entropy estimation methods in App. \ref{app:EntropyEstimation}. These include an analytic estimation of the entropy, subsystem-based estimation and variationally disentangling the state. For our purposes, the most practical of these approaches is the scaled subsystem entropy, which we detail here.

Hamiltonians of many-body systems are typically translation invariant or possess some spatial symmetry. These symmetries allow us to approximate the entropy of the full system by computing the entropy of a subsystem that is accessible to tomography or classical simulation and then rescaling to the larger system size.


Let us consider a simple example of a one-dimensional system arranged on a ring. The scaled entropy of the subsystem can be evaluated for a subsystem of $n_a$ qubits and rescaled to $n$ qubits. 
\begin{equation}
    S_{n_a}(\bm{\theta,\lambda}) = \frac{n_a}{n} S(\mathcal{A}_{\bm{\theta,\lambda}}(\rho_0)|_{n_a}),
\end{equation}
with $\mathcal{A}_{\bm{\theta,\lambda}}(\rho_0)|_{n_a}$ is the ansatz for $n_a$ qubits. 

Now consider a specific ansatz $\mathcal{A}_{\bm{\theta,\lambda}}$ that consists of unitaries interleaved with single-qubit Pauli channels. 
For high-temperature states, the entropy approximation is easily justified. Above a certain temperature, the Gibbs state is well approximated by a mixture of product states, showing zero entanglement even at short range \cite{bakshi_high-temperature_2024}. Consequently, the entropy approximation is exact in the infinite temperature limit. 
From the perspective of the circuit preparing the state, large temperatures correspond to a high Pauli error rate. In this regime, the output of the circuit converges exponentially to the fully mixed state as the number of layers increases~\cite{wang_noise-induced_2021}. For the fully mixed state, the entropy approximation is exact.

For small Pauli error rates justifying the entropy approximation is more involved. We show that for random circuit instances, the error of the approximation decays exponentially with the size of the subsystem used for the approximation. The output of a random quantum circuit interleaved with Pauli channels is well described by a pure state $\ket{\Psi}$ and uniform noise \cite{boixo_Characterizing_2018,dalzell_random_2021}:
\begin{equation}\label{eq:stateBrandaoa}
    \rho = \alpha \ketbra{\Psi}{\Psi} + (1-\alpha)\frac{\mathbb{I}}{2^n}, \quad \alpha = e^{-\lambda n m}.
\end{equation}
Here $\lambda$ is the single-qubit Pauli error rate, $n$ the number of qubits and $m$ the number of layers. 
This approximation is meaningful for high fidelity with the pure state,
$\lambda \sqrt{m n}\ll 1$, logarithmic depth $m \geq \Omega(\log(n))$ and small error rates $\lambda^{-1} \leq \tilde{\Omega}(n)$.
The entropy of this state evaluates to 
\begin{align}
S = & -\left(\alpha + \frac{1 - \alpha}{2^n}\right) \log\left(\alpha + \frac{1 - \alpha}{2^n}\right) \nonumber \\
&- \frac{1 - \alpha}{2^n}(2^n - 1) \log\left(\frac{1 - \alpha}{2^n}\right).
\end{align}
Using the expression for the entropy for system sizes $n$ and $n_a$ we approximate the absolute error of the entropy approximation
\begin{align}\label{eq:EntropyErrorMain}
    |S_n - S_{n_a}| \approx e^{-n_a \lambda m} n \log(2).
\end{align}
The relative error scales as 
\begin{align}\label{eq:RelativeEntropyErrorMain}
    \left|\frac{S_n - S_{n_a}}{S_n}\right| \approx e^{-n_a \lambda m}.
\end{align}
The error of the approximation decreases exponentially with the size of the subsystem used in the approximation allowing a subsystem of logarithmic size to estimate the entropy of a large system.

\begin{figure}
    \centering
    \includegraphics{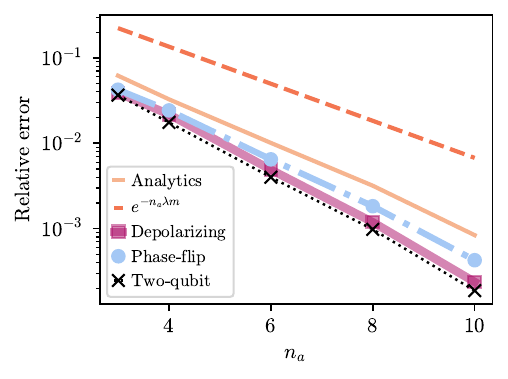}
    \caption{Relative error of the entropy approximation. The analytic expression for the error of the entropy approximation (Eq. \eqref{eq:RelativeEntropyErrorMain}) compared to numeric simulations of random circuits. The approximation error is calculated for different subsystem sizes $n_a$ with respect to the exact entropy for $n=12$ qubits. The parameters for the unitary layers are sampled from the interval $\left[0,2\pi \right)$. The Pauli error rate for the depolarizing and phase-flip channel is $0.05$. The multi-qubit nonunitary operations allow at maximum $7\%$ of the population to decay. The approximation error decays exponentially in the size used for the approximation and agrees well with the analytic expression.}
    \label{fig:RelativeEntropyErrora}
\end{figure}

In Fig.~\ref{fig:RelativeEntropyErrora} we compare the analytic expression in Eq. \eqref{eq:RelativeEntropyErrorMain}, to the expectation value of the simulated circuits. We plot the approximation error of the scaled subsystem entropy to the exact entropy evaluated for 12 qubits. The circuit consists of 10 layers of $ZZ$ and $X$ rotations interleaved with nonunitary components. 
We show depolarizing and phase-flip channels with the parameter $\lambda = 0.05$. The two-qubit channel is generated by a Lindbladian evolution (App. \ref{ap:jump_Heis}). The detunings for the two-qubit operations are sampled from the interval $\left[0,1 \right)$ and the evolution time is sampled from the interval $\left[ 0,10^3 \right)$ allowing at most $7\%$ of the population of a computational basis state to decay. 
Introducing two-qubit nonunital operations goes beyond the assumptions made in Ref. \cite{dalzell_random_2021} making the analytic approximation of the error no longer applicable. However, the numerics show similar behavior.
For all ans\"atze, the approximation error decays exponentially in system size and is in good agreement with the analytic expression. 
The analytic expression is derived for single-qubit Pauli noise but qualitatively also holds for multi-qubit nonunitary operations.

Although the error is small for average problem instances, optimizing with the approximate entropy can be challenging. Average instances well approximate the behavior of the circuit when it is randomly initialized. During optimization, the optimizer determines which parameters are sampled. The optimizer that minimizes the free energy can reduce its cost by increasing the entropy of the subsystem, while the entropy of the entire system is not changed. As a consequence, the global minimum of the cost landscape does not produce the target Gibbs state. To combat this, we introduce a regularization term to our cost function. We multiply it by $1-|\Delta S|$ where $\Delta S = S_{n_a}-S_{n_b}$ is the difference in entropy calculated with two smaller instances of the circuit $n_a$ and $n_b$. In practice, we choose $n_a = 3$ and $n_b = 4$. The total cost function is
\begin{equation}\label{eq:RegularizedCost}
    (1-|\Delta S|) (E-TS_{n_a}).
\end{equation}
The entropy used to calculate the free energy is taken to be that of the smaller subsystem $n_a$.

Translation invariance is integral to the scaled entropy approximation. In many-body quantum simulation problems translation invariance is mostly given, making this method for entropy estimation particularly useful in this context. However, the promising application for Gibbs state preparation, quantum Boltzmann machines (QBM), considers general Hamiltonians \cite{amin_Quantum_2018,zoufal_variational_2021}. In the training of a QBM the Hamiltonian parameters are optimized such that the distribution produced by the Gibbs state of the Hamiltonian matches a target distribution. For this task, translation symmetry is not given. To estimate the entropy one can split the quantum system into subsystems that are tractable for classical simulation, evaluate their entropy, and add it to estimate the entropy of the whole system. We consider further methods for entropy estimation in App. \ref{app:EntropyEstimation}.

\section{Weak and strong symmetries}
Symmetries greatly improve our understanding of the laws of nature, providing fundamental insights into the behavior of physical systems and conservation laws. In quantum simulation, understanding and leveraging these symmetries allows for more efficient modeling of complex quantum systems \cite{chen_boosting_2024}. For VQTs, symmetries improve trainability and aid in finding appropriate nonunitary operations. Here we only introduce the most relevant concepts; for a more comprehensive introduction, see App. \ref{app:Symmetries}.

Consider a model Hamiltonian $H$ with symmetry group $G$ that has representation $R_g$ for $g$ in $G$. Then, per definition,
\begin{equation}
    \left[R_g,H\right] = 0.
\end{equation}
It directly follows that Gibbs state share the symmetry of the model Hamiltonian
\begin{equation}
    \left[R_g,\rho_G\right] \propto \left[R_g,\sum_{n=0}^\infty \frac{H^n}{n!}\right] = \sum_{n=0}^\infty \frac{1}{n!}\left[R_g,H^n\right]=0.
\end{equation} 
To simplify the notation, we absorb $-\beta$ in the Hamiltonian.
We aim to prepare this state by means of a quantum circuit consisting of an interleaved structure of unitary and nonunitary layers as introduced in Sec. \ref{sec:ansatz}.

Symmetries of unitary variational quantum algorithms have been extensively studied \cite{zheng_sncqa_2023,sauvage_building_2024,gard_Efficient_2020,park_efficient_2021,meyer_exploiting_2023,larocca_group-invariant_2022}. In the context of state preparation, the most common approach is to initialize the problem in the symmetry sector of the target state and then constrain the search space by only considering unitaries that respect the symmetries \cite{gard_Efficient_2020}. The advantage of this approach is improved trainability due to the reduced size of the state space explored by the optimizer \cite{sauvage_building_2024,schatzki_theoretical_2024, meyer_exploiting_2023}. In quantum machine learning where encoding the right bias into the model is essential and one wants to avoid overfitting, this advantage is amplified \cite{larocca_group-invariant_2022, muser_provable_2024}. Normally, in state preparation overfitting does not present an issue; we, however, train with an approximation of the entropy; therefore, an optimizer could overfit by exploiting errors in the approximation. In this regard, a symmetric ansatz is more constrained.

On the other hand, it may be desirable to break symmetries if the initial state is not in the same symmetry sector as the target state \cite{park_efficient_2021}. In the problems we consider, we initialize the problem in the symmetry sector of the ground state. We then apply unitary layers that respect the symmetry 
\begin{equation}
    R_g U_\theta R_g^\dag = U_\theta, \quad \forall g,\theta.
\end{equation}

 \begin{figure}
     \centering
     \includegraphics[width = \columnwidth]{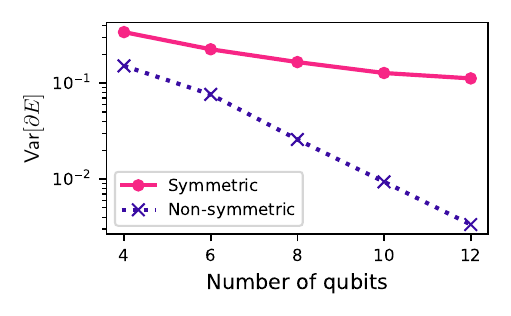}
     \caption{The effect of symmetries on gradient variances. We consider $\prod X_i$ symmetry and compare a symmetric ansatz that is built from $ZZ$ and $X$ rotations with an ansatz that is built from $ZZ$, $X$ and $Z$ rotations. Independent of the number of qubits this circuit primitive is repeated 40 times. The graph compares the estimated variance of the gradient when measuring a $ZZ$ Pauli string on the first two qubits. 
     The variance of the gradient decays exponentially in system size, which by Chebyshev's inequality can be related to the gradient itself. The gradient in the symmetric case decays slower but still exponentially. This can be attributed to the space explored by the ansatz scaling exponentially with the size of the system but at a smaller rate.
     }
     \label{fig:GradiantVariances}
 \end{figure}
The advantages of symmetric unitary layers can be understood from Fig. \ref{fig:GradiantVariances}. We plot the variance of the gradient of the energy term for different system sizes. To isolate the effect of the unitary layers, the non-unitary elements are set to act as the identity, making the setting equivalent to a unitary parameterized circuit \cite{meyer_exploiting_2023}. We compare two ans\"atze: A symmetric ansatz built from $ZZ$ and $X$ rotations and another ansatz constructed from $ZZ$, $X$, and $Z$ rotations. These layers are repeated 40 times, regardless of the number of qubits. We focus on the estimated gradient when measuring a $ZZ$ Pauli string on the first two qubits. The variance of the gradient decays exponentially with system size, which, according to Chebyshev's inequality, can be related to the gradient itself. In the symmetric case, the gradient decays more slowly, but still exponentially. This behavior can be attributed to the exponential scaling of the ansatz's space with the system size, albeit at a reduced rate.

\begin{figure}
    \centering
    \includegraphics{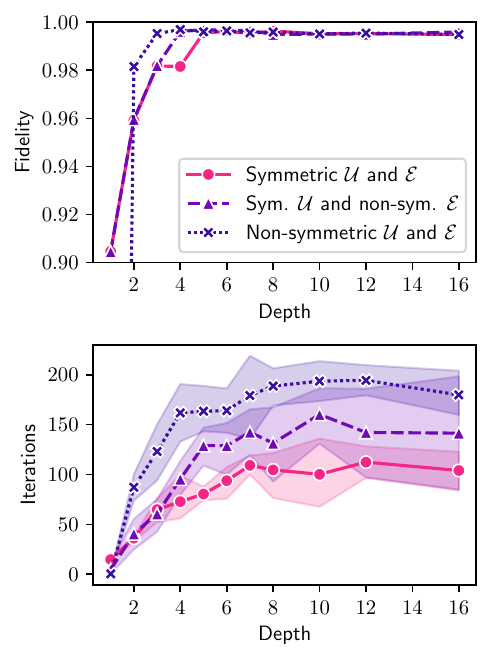}
    \caption{The effect of unitary symmetries on model training. Six qubits TFIM trained with the exact entropy $\beta = 0.75$ (a) Fidelity of the output state with the target state for the run that converged to the lowest cost-free energy. (b) Iterations needed to converge. Restricting the ansatz to be symmetric reduces the number of training steps needed to converge.}
    \label{fig:trainability_symmetry}
\end{figure}

The practical consequences in optimization can be seen in Fig. \ref{fig:trainability_symmetry} where we compare the optimization iterations needed up to convergence for symmetric and non-symmetric ans\"atze. Choosing a symmetric ansatz halves the optimization iterations needed for convergence, which is in agreement with the reduced state space the symmetric ansatz explores. In practice, such improvements allow for larger systems to be studied.

Quantum channels and dissipative evolution are more general than unitary dynamics and therefore also have a richer interplay with symmetries. We introduce two symmetry definitions \cite{buca_note_2012,de_groot_symmetry_2022}:
\begin{definition}[Weak symmetry]
    A channel $\mathcal{E}$ satisfies the \textit{weak symmetry condition} if it commutes with the channel representing the symmetry; $\mathcal{R}_g(\rho) = R_g \rho R_g^\dag$.
    Specifically, for all $g$ in $G$:
    \begin{equation}
        \mathcal{R}_g \circ \mathcal{E} \circ \mathcal{R}^\dag_g = \mathcal{E}.
    \end{equation}
\end{definition}

The Gibbs state commutes with the symmetry ($\left[ R,\rho_{G}\right] = 0$). Respecting the weak symmetry ensures that the output state of the VQT commutes with the Hamiltonian. Assume initial state $\rho_0$ commutes ($\left[ R,\rho_0 \right]=0$) then
\begin{align}
    R \mathcal{E}(\rho_0 ) &= R \mathcal{E}(\rho_0 ) R^\dag R \nonumber\\
    &= \mathcal{E}(R\rho_0 R^\dag )  R \nonumber\\
    &= \mathcal{E}(\rho_0 )  R,
\end{align}
the output of the VQT also commutes with the symmetry. Restricting the channels to be weakly symmetric ensures that the output state of the ansatz has the correct symmetry properties. 

A quantum channel can respect symmetry in a stricter sense: 
\begin{definition}[Strong symmetry]
    A Kraus channel with Kraus operators $\{K_i\}$ meets the strong symmetry condition if
    \begin{equation}
        R_g K_i R_g^\dag = e^{i\theta(g)} K_i, \forall i,g.
    \end{equation}
    for a phase function $\theta: G \mapsto \mathbb{R}$.
\end{definition}
We will analyze this from the view of symmetry sectors. 
The projector onto the symmetry sector of irreducible representation $\alpha$ can be expressed as
\begin{equation}
    \Pi_\alpha = \frac{1}{|G|} \sum_g \chi_\alpha (g) R_g.
\end{equation}
Here $\chi_\alpha(g)=\Tr\left[\gamma_\alpha(g)\right]$ is the character of irrep $\gamma_\alpha$. 
By construction, the projectors $\Pi_\alpha $ are mutually orthogonal and commute with the Hamiltonian. The image of each projector is referred to as a \textit{symmetry sector} \cite{misguich_detecting_2007}. 

The overlap of state $\rho$ with symmetry sector $\alpha$ is
\begin{equation}
    p_\alpha  = \Tr\left[\Pi_\alpha \rho\right].
\end{equation}
A strongly symmetric channel leaves the set $\{ p_\alpha \}$ invariant (see App. \ref{sec:symsec}). For an ansatz consisting of symmetric unitary layers and strongly symmetric channels the set $\{ p_\alpha \}$ will be the same as that of the output state.

Individual terms in the mixture that comprises the Gibbs state can be in different symmetry sectors and the set of $\{ p_\alpha \}$ will likely differ from that of the desired Gibbs state. The simplest way to see this is to consider the infinite temperature state: this can be expressed as the equal mixture of all basis states in any basis and therefore has nonzero overlap with all symmetry sectors. As a consequence, it is essential that the channel can break the strong symmetry. This allows different terms in the mixture to occupy all symmetry sectors.

As an example, consider a model with symmetry $R= X^{\otimes n}.$
We compare symmetric unitary layers combined with weakly and strongly symmetric channels.
We consider the bitflip channel with Kraus operators proportional to $\{ \mathbb{I}, X\}$ both Kraus operators commute with symmetry $R_X = X$ and therefore the channel is strongly symmetric. As a consequence, the channel is also weakly symmetric.

The second channel we consider is the phaseflip channel with Kraus operators proportional to $\{ \mathbb{I} , Z \}$. The $Z$ operator does not commute with the symmetry, $\left[ X,Z\right] \neq 0$, and the channel is not strongly symmetric. Weak symmetry can be verified quickly using the property $ZX = -XZ$. 

The resulting fidelities with the target Gibbs state are plotted in Fig \ref{fig:Story_double} (c) for different inverse temperatures, $\beta$. On the right, at high inverse temperatures, the ground state is prepared. This is analogous to the standard VQE setting. All ans\"atze can prepare the ground state with high fidelity as the initial state lies in the same symmetry sector as the ground state. With decreasing inverse temperature (increasing temperature), states from other symmetry sectors appear in the mixed target state. Through an odd number of $Z$ operations, the phaseflip channel can switch symmetry sectors while the bitflip channel is confined to the initial symmetry sector. Therefore, the bitflip channel can prepare high or infinite temperature states with at most fidelity of 0.5 as can be seen from Fig. \ref{fig:Story_double}.

In practice, the advantages of using symmetries to constrain the ansatz are shown in Fig. \ref{fig:trainability_symmetry}. We aim to prepare the Gibbs state of the TFIM at $\beta = 0.75$ for six qubits and compare ans\"atze with symmetric and nonsymmetric unitaries and channels. All ans\"atze reach approximately the same fidelity for more than five layers. The number of optimization steps it takes to reach this fidelity differs. Restricting the unitaries to be symmetric improves the trainability, and restricting the channels provides a further advantage. Together, these restrictions make for half the number of optimization steps.

The concepts of weak and strongly symmetric channels improve our understanding of thermal state preparation. Practically, the number of training steps is reduced by respecting the symmetry weakly as a consequence of the smaller state space the ansatz explores. Respecting the symmetry strongly constrains the ansatz too much. Finally, when engineering nonunitary operations symmetry considerations are a useful aid that can restrict the set of possible channels that have to be considered. 
\section{Thermal states of quantum spin models}\label{sec:spinthermal}
Quantum spin chains are rich models that exhibit many interesting physical phenomena. There is a zoo of quantum spin chains that have been studied going back almost a century. We select a few paradigmatic models to demonstrate the potential of our algorithm for preparing thermal states in a wide range of models. The numerical experiments in this section are performed with 6 qubits arranged on a ring. For clarity of the presentation, we neglect the fluctuations due to the inefficient optimization procedure by running with multiple, randomly chosen, initializations and using the run that converged to the minimal cost function.

To assess our algorithm's performance, we need a metric that is independent of variables like the model Hamiltonian, system size, or target temperature. One promising candidate is the cost function utilized during training: the free energy. While it's convenient for training purposes, its dependence on the Hamiltonian spectrum and temperature renders it inadequate for comparisons.

The most widely employed metric for comparing two mixed states is the Uhlmann and Jozsa fidelity, 
\begin{equation}
		\mathcal{F} (\rho ,\sigma )= \max_{\ket{\psi},\ket{\phi}} |\bra{\psi} \ket{\phi}|^2 =  \left( \mathrm{Tr} \sqrt {{\sqrt {\rho}} \sigma {\sqrt {\rho }}} \right) ^{2},
\label{eq:fidelity_UJ}
\end{equation}
which takes the maximal transition probability between the purifications $\psi$ and $\phi$ of a pair of density matrices $\rho$ and $\sigma$. Other metrics such as the trace distance, the Kolmogorov complexity, and the quantum cross-entropy can also be employed to quantify how close the prepared state is to the desired one. Previous studies have found that these different metrics show qualitatively similar behavior, and therefore we focus on fidelity to judge the success of our state preparation \cite{selisko_extending_2024}.

It is important to note that performance metrics that rely on explicit representations of the target Gibbs state, such as fidelity, are not scalable. For the algorithm to provide a quantum advantage, the classical representation of the state has to be intractable. Without this representation of the state, the fidelity cannot be calculated. Verifying the accuracy of the state preparation for large systems requires methods such as Hamiltonian tomography and is an active area of research \cite{lifshitz_practical_2023,gu_practical_2024}.

\subsection{Ising model}
\begin{figure}
    \centering
    \includegraphics[width = \columnwidth ]{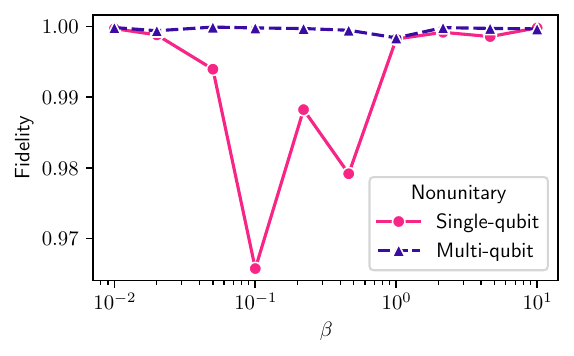}
    \caption{Ising model. We plot the fidelity with the ideal target state for different inverse temperatures $\beta$. We compare single-qubit bitflip channels with the addition of two-qubit projectors. The projectors allow the thermalizer to prepare the desired target state with fidelity$>0.99$ at all temperatures. }
    \label{fig:Ising}
\end{figure}

The Ising model is described by Hamiltonian
\begin{align}
    H = -J \sum_k Z_k Z_{k+1}  - g \sum_k Z_k. 
\end{align}
Being non-frustrated its ground state can be prepared by projecting onto the ground states of the individual $Z_k Z_{k+1}$ terms \cite{kraus_Preparation_2008}. We compare two ans\"atze that use the same unitary $ZZ$ and $X$ rotations. The nonunitary components differ: One relies on the single-qubit bitflip channel with Kraus operators $\{\sqrt{1-p}\mathbb{I},\sqrt{p}X\}$. The multi-qubit ansatz uses two-qubit nonunitary operations that probabilistically apply projectors onto the ground states of the individual $Z_kZ_{k+1}$ terms.
We compare single-qubit nonunitary operations with projectors in Fig. \ref{fig:Ising} for $J=g=1$. 
From here we see that multi-qubit nonunitary operations greatly improve fidelity, achieving near-perfect state preparation.

\subsection{Transverse-field Ising model}
The transverse-field Ising model is the quantum version of the classical Ising model, featuring non-commuting $ZZ$ and $X$ terms. This makes it one of the simplest frustrated models and a popular benchmark for quantum state preparation and time evolution algorithms \cite{scholl_Quantum_2021,kim_Evidence_2023}. The model's Hamiltonian is expressed as:
\begin{align}
    H = -J \sum_k Z_k Z_{k+1}  - g \sum_k X_k. 
\end{align}
We focus on the thermal states at the critical point $J = g = 1$ \cite{sachdev_quantum_2011}. 

We initialize the system in the ground state of the transverse field, $\ket{+}^{\otimes n}$, and evolve with the mixing Hamiltonian $H_m = -\sum_k X_k$ and the coupling Hamiltonian $H_c= - \sum_k Z_k Z_{k+1}$ \cite{moll_Quantum_2018}. 
The phaseflip channel consisting of Kraus operators $\{\sqrt{1-p} \mathbb{I}, \sqrt{p} Z\}$ is a good choice starting point for a nonunitary ingredient as it is weakly symmetric and therefore suitable from a symmetry standpoint. 
Furthermore, it maps between eigenstates of the transverse field term of the model. Specifically, the operation $Z$ on any qubit maps this ground state to the first excited state, making $Z$ the ideal heating operator in the early stages of the evolution. 

\begin{figure}
    \centering
    \includegraphics[width = \columnwidth ]{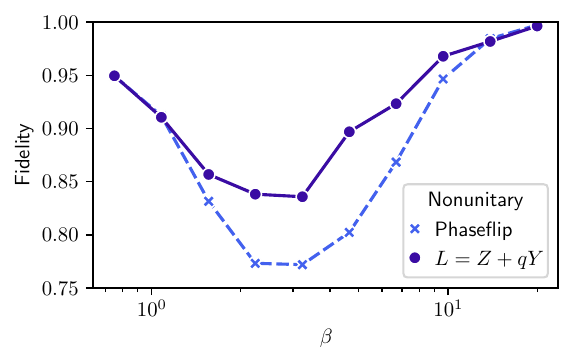}
    \caption{Transverse-field Ising model. Thermal states of the transverse-field Ising model at its critical point. We compare an ansatz that uses the phaseflip channel to channel with an ansatz that uses a channel with jump operators of the form $L = Z+qY$. Both ans\"atze were trained with the approximate entropy.}
    \label{fig:TFIM}
\end{figure}

Through evolution with $H_m$ and $H_c$ the state, in the absence of nonunitary elements, evolves towards the ground state of the model Hamiltonian. As the state and system evolve, the jumps should also be adjusted. According to App. \ref{ap:Connection} we adjust the nonunitary evolution to Lindbladian evolution generated by
\begin{align}\label{eq:jump_TFIM}
    L = Z + q Y. 
\end{align}
How we would realize such a jump in a physical system is described in App. \ref{ap:jump_TFIM_experiment}.
In Fig. \ref{fig:TFIM}, we plot the fidelity with the ideal thermal states for different inverse temperatures. We find that the channel generated by Eq. \eqref{eq:jump_TFIM} significantly improves the performance of the algorithm. 
\subsection{Heisenberg model with a transverse field}
The Heisenberg model is described by the Hamiltonian:
\begin{align}
    H = -\sum_k \left(X_k X_{k+1} + Y_k Y_{k+1} + Z_k Z_{k+1} \right)- \Delta \sum_k X_k  .
\end{align}
We prepare the thermal states of the model for $\Delta = 1$. 

A popular ansatz for the Heisenberg model consists of initializing in the ground state for pairs of qubits and then evolving with even and odd terms in an alternating fashion. We found that in the presence of a transverse field, the ansatz for the transverse-field Ising model offers better performance. 
Consequently, the phaseflip channel is a good starting point for the nonunitary contribution.

Thermal states can be prepared with reasonable fidelities with an ansatz that uses single-qubit noise. At intermediate temperatures i.e. close to the ground state but nonpure, the fidelity with the target state is low. At these temperatures, multi-qubit nonunitary operations could improve the performance. 

There is much freedom in the choice of multi-qubit nonunitary operations. The symmetries of the model can be used to narrow down the choice of appropriate operations. To this end, the couplings should be invariant under inversion of all spins; $R = \prod_j X_j$. We engineer effective two-body decays in a pair of three-level systems coupled to two oscillator modes. By adjusting the system parameters, we can control the relative strength of the jumps towards ferromagnetic and antiferromagnetic states, allowing for precise tuning of the dissipator.
\begin{figure}
    \centering
    \includegraphics[width=\columnwidth]{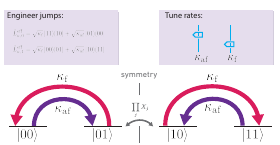}
    \caption{Nonunitary multi-qubit gate operation. Two-body jump operators employed in the VQT for the Heisenberg model. We consider jumps that align or antialign neighboring spins. The action of the symmetry swaps the effective jump operators making the channel they generate weakly symmetric. The decay rates $\kappa_\text{f}$ and $\kappa_\text{af}$ along with the evolution time parameterize the action of the channel. }
    \label{fig:MultiQubitNonunitaryOperation}
\end{figure}
We engineer the effective jump operators of the form
\begin{align}
    L_{\kappa,0} ^\text{eff} &= \sqrt{\kappa_\text{f}}\ketbra{11}{10} + \sqrt{\kappa_\text{af}}\ketbra{01}{00},\\
    L_{\kappa,1}^\text{eff} &= \sqrt{\kappa_\text{f}}\ketbra{00}{01} + \sqrt{\kappa_\text{af}}\ketbra{10}{11}.
\end{align}
The evolution generated by these jump operators satisfies the symmetry weakly as $R L_{\kappa,0} ^\text{eff} R= L_{\kappa,1} ^\text{eff}$ and therefore $\mathcal{R} \circ \mathcal{L} \circ \mathcal{R}^\dag = \mathcal{L}$. The effective Lindbladian breaks the strong symmetry as the individual jump operators do not commute with the action of the symmetry. 
We show the effective jump operators in Fig. \ref{fig:MultiQubitNonunitaryOperation} and in App. \ref{ap:jump_Heis} their physical implementation is discussed.

We evaluate the performance of our algorithm in the XXZ Heisenberg model with a transverse field. The use of multi-qubit nonunitary operations significantly increases the performance of the algorithm. This increase in performance cannot be attained by a higher depth ansatz that uses single-qubit noise as ancillas would be required to emulate the two-qubit dissipators. 
Multi-qubit nonunitary operations cannot be emulated via single-qubit nonunitary operations along with unitary operations \cite{zapusek_nonunitary_2023}. 
\begin{figure}
    \centering
    \includegraphics[width = \columnwidth]{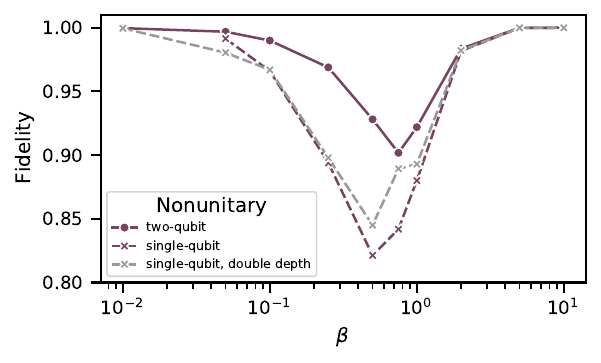}
    \caption{Heisenberg model. Comparison of the fidelity with the Gibbs state in the Heisenberg model with a transverse field for six qubits on a ring, using an ansatz with and without multi-qubit nonunitary operations. The fidelity after convergence is plotted for different inverse temperatures $\beta$. The results show that nonunitary multi-qubit operations improve the accuracy significantly. }
    \label{fig:MultiQubitChannels}
\end{figure}

\section{Conclusion and Outlook}\label{sec:Conclusion}
We have investigated the preparation of thermal states by a quantum-classical ansatz that uses both nonunitary and unitary layers. Through the use of symmetries of the problem Hamiltonian, we simplify the optimization problem. 
Going further, we apply the concept of weak and strong symmetries for the first time in the context of parameterized quantum algorithms. For the preparation of thermal states, we find strong symmetries to be too restrictive, while weak symmetries prove to be a useful tool. They improve the understanding of thermal state preparation and facilitate the design of multi-qubit nonunitary operations.

We propose practical implementations of the nonunitary circuit elements, which can be realized on present-day trapped ion setups without overhead in physical qubits \cite{van_mourik_experimental_2024}. 
By connecting our near-term algorithm and the Gibbs sampler from Ref. \cite{chen_efficient_2023}, we demonstrate that at high temperatures our dynamics can reproduce those of the Gibbs sampler, which converges to the Gibbs state as its steady state. Building on this connection, we design nonunitary operations that enhance the preparation of the transverse-field Ising model. The use of multi-qubit nonunitary operations significantly improves fidelity with the ideal state for both the Ising model and the Heisenberg model with a transverse field. 
These improvements for paradigmatic models underscore the broad applicability of our approach and pave the way for new avenues in quantum simulation of open systems, making VQT a viable tool for materials simulation, quantum chemistry, and algorithmic applications.

In our numerical experiments, fidelity is lowest at intermediate temperatures, a trend observed in other Gibbs samplers as well \cite{foldager_noise-assisted_2022,consiglio_variational_2024,selisko_extending_2024}. This raises a deeper question: Why are intermediate temperatures hardest to prepare? Ground state preparation is QMA-complete, while high-temperature Gibbs states can be well-approximated with shallow circuits \cite{bakshi_high-temperature_2024}. Since our models allow for good ground-state preparation and high-temperature states are always easy to prepare, the lowest fidelity must occur at intermediate temperatures. However, this is not a fundamental limitation but rather a consequence of our choice of models and ans\"atze.
It is of interest to further investigate the universality of the design principles we present. To this end, one could consider further Hamiltonians and spatial geometries. One could even extend this to the generation of generalized Gibbs ensembles \cite{reiter_engineering_2021}.

Investigating algorithms that utilize dissipation to enhance robustness against noise has become a topic of growing interest \cite{harrington_engineered_2022}. In this context, we present primitives that may find application in such noise-resistant algorithms.
Specifically, the protection of lattice gauge theories against errors shows promise \cite{schmale_Stabilizing_2024,wauters_symmetry-protection_2024}. Here the concepts of weak and strong symmetries could be used to further the understanding of the correction operation and the scalability of the method. Furthermore, the design principles we use to engineer the nonunitary operations could be used to design stabilizing operations.

The data to support the findings of this study is available at Ref. \cite{VQT_data}.
\begin{acknowledgments}
We wish to acknowledge insightful discussions with Ivan Rojkov and Zoe Holmes. F.R. and E.Z. acknowledge funding from the Swiss National Science Foundation (Ambizione grant no. PZ00P2 186040) and the ETH Research Grant ETH-49 20-2. K.K. acknowledges financial support from the QSIT/Quantum Center at ETH Zurich and funding from the QuantumReady project (FFG 896217) within Quantum Austria (managed by the FFG).
\end{acknowledgments}

\appendix

\section{Approximations of Entropies}\label{app:EntropyEstimation}
Evaluating the entropy of the system is essential to the success of our algorithm. Here we derive, discuss and benchmark different approaches to estimate the entropy. 

Measuring the von Neumann entropy;
\begin{equation}
    S(\rho) = - \Tr[\rho \log(\rho)],
\end{equation}
is hard as it is a nonlinear function of the state and therefore not a quantum observable. In general, it requires knowledge of the $2^n$ eigenvalues of the density matrix.
To calculate it, one needs access to the system's density matrix which requires full-state tomography which is infeasible for moderately sized systems. Even though calculating the entropy of a general system is resource-expensive, our system is produced by a specific quantum circuit. We can use our knowledge of the circuit to approximate the entropy of the output state.
\subsection{Scaled Subsystem Entropy}
In the main text, we introduce a scaled entropy approximation for translation-invariant systems. The entropy of a subsystem of size $n_a$ is computed and rescaled to approximate the entropy of the full system. For further details, refer to Sec. \ref{sec:cost} of the main text.

\subsection{Analytic Entropy approximation}\label{ap:analyticderivation}
 The authors of Ref.~\cite{foldager_noise-assisted_2022} derive an analytic approximation of the entropy of an ansatz that interleaves $m$ unitary layers with $m$ layers of depolarizing channels. The authors calculate the entropy of the state produced by an ansatz that first applies all $m$ layers of the depolarizing channel followed by $m$ unitary layers. 
 The approximation is motivated by the fact that the depolarizing channel commutes with all products of single-qubit unitaries. As a consequence, the entropy of the state produced by this ansatz is close to the entropy of the ansatz that is used in practice.
The definition of the depolarizing channel is:
\begin{equation}
    \mathcal{D}_\lambda(\rho) = (1-\lambda)\rho -\frac{\lambda}{2} \mathbb{I}.
\end{equation}
The composition of $m$ depolarizing channels can be written as a single depolarizing channel
\begin{equation}
    \mathcal{D}_\lambda \circ \mathcal{D}_\lambda  \circ \ldots \circ \mathcal{D}_\lambda  = \mathcal{D}(\Lambda),
\end{equation}
with parameter $\Lambda=1-(1-\lambda)^m$. 

For a single-qubit, the output state is 
\begin{equation}
\mathcal{D}_\Lambda(\ketbra{+}{+}) = (1-\frac{\Lambda}{2})\ketbra{+}{+} +\frac{\Lambda}{2}\ketbra{-}{-}.
\end{equation}
From this expression, the entropy can be directly determined as $S = (1-\Lambda/2)\log(1-\Lambda/2) + \Lambda \log(\Lambda/2)$. 
In our approximation the system stays in a product state therefore, the entropy of the whole system is the entropy multiplied by the number of qubits:
\begin{equation}
    S \approx -n \left( \left(1- \frac{\Lambda}{2} \right)\log(1-\frac{\Lambda}{2}
)  +\frac{\Lambda}{2} \log(\frac{\Lambda}{2})\right).
\label{entropy_approximation}
\end{equation}
Since this expression is analytical, it can be evaluated efficiently and does not require any measurements. 
However, in our circuit, the nonunitary elements are interleaved with the unitary ones. Hence, Eq.~\ref{entropy_approximation} is only an \textit{approximation} to the entropy of the state produced from our quantum circuit. Moreover, it is only valid for the highly symmetric depolarizing channel.

\subsection{Variational Entropy Estimation}
Evaluating the entropy of a product state is as simple as
\begin{equation}
    S(\rho_{1} \otimes \rho_2 \otimes \dots \otimes \rho_n) = S(\rho_1)+S(\rho_2)+\dots+S(\rho_n).
\end{equation}
Here $\rho_j$ signifies the partial trace over all subsystems but the $j$-th one. For a general state, the entropy is upper bounded by the sum of entropies of subsystems
\begin{equation}
    S(\rho) \leq S(\rho_1)+S(\rho_2)+\dots+S(\rho_n).
\end{equation}
Furthermore, the entropy of a quantum state is not changed by unitary transformations. From this identity stems the idea to transform the state $\rho$ as close to a product state and then upper bound its entropy through the entropies of subsystems.

The unitary that removes quantum correlations is found variationally by minimizing the cost function
\begin{equation}
    C(U_{\phi} \rho U_{\phi}^\dag ) = S(\tilde{\rho}_1^{\phi})+S(\tilde{\rho}_2^{\phi})+\dots+S(\tilde{\rho}_n^{\phi}).
\end{equation}
Here $ \tilde{\rho}_j^{\phi} = \Tr_j [ U_{\phi} \rho U_{\phi}^\dag ]$, the $j$-th reduced density matrix after the unitary was applied. Learning this unitary is analogous to the quantum state learning problem discussed in Ref. \cite{verdon_quantum_2019}. In fact, by solving the quantum state learning problem one obtains an estimate of the entropy of the state.

The preparation algorithm for Gibbs states now contains two optimization loops. The first loop minimizes the system's free energy. The second loop calculates the entropy. Although it seems costly, the second loop can converge quickly. We know the circuit that creates the state whose entropy we want to estimate. While we cannot invert the entire circuit, we can invert the unitary parts. For this, we choose an ansatz and initialization that inverts the unitary portion of the preparation ansatz.

Training with an upper bound of the entropy provides additional challenges. The optimizer that minimizes the free energy can decrease its cost by choosing states for which this upper bound is not tight. Ultimately, this leads to the optimizer converging to a state far from the target Gibbs state. The bound can be improved by considering larger subsystems. Using the subadditivity twice we find
\begin{equation}
    S(\rho_{1234}\leq S(\rho_{12}) + S(\rho_{34}) \leq S(\rho_{1}) + S(\rho_{2})  + S(\rho_{3}) + S(\rho_{4}). 
\end{equation}
Using this second estimate, we introduce a regularization term much like in the scaled subsystem entropy (Eq. \eqref{eq:RegularizedCost}).
Now, $S_A$ is the variationally estimated entropy using single-qubit subsystems, while $S_B$ uses two-qubit subsystems.  
\subsection{Comparison for translation invariant system}
We consider a system with translation invariance. This allows us to benchmark the scaled subsystem entropy against the variational entropy estimation. We consider the Heisenberg model with a transverse field
\begin{equation}
    H = \sum_k  X_k  X_{k+1}+ Y_j  Y_{k+1}+ Z_k  Z_{k+1} + \sum_k  X_k.
\end{equation}
We initialize the ansatz in the symmetry sector of the ground state. The unitary part of the ansatz $\mathcal{A}_{\bm{\theta},\bm{\lambda}}$ respects the symmetry, while the phase-flip noise we consider breaks the strong symmetry of the ansatz. 
\begin{figure*}
    \centering
    \includegraphics{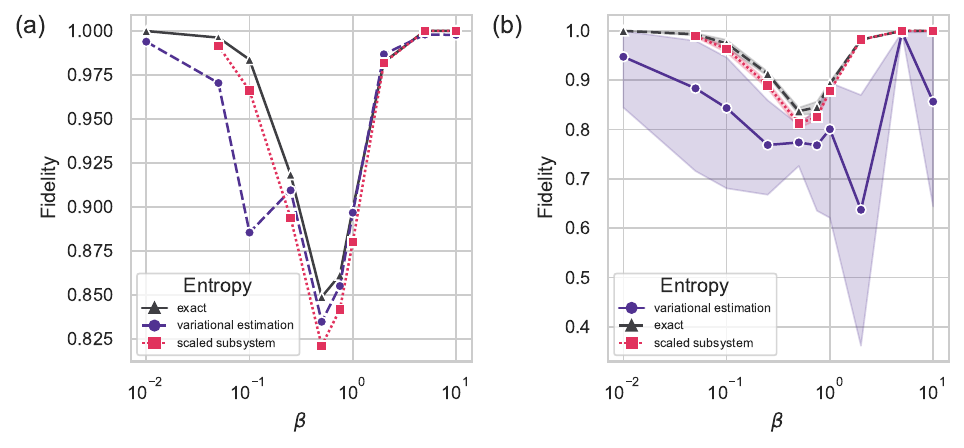}
    \caption{Comparison of entropy estimation methods. (a) Fidelity with the target state for the lowest final cost. The circuit was initialized randomly for 10 different seeds. (b) Average converged fidelity with the target state. The error bars represent the standard deviation. The large variance for the variational entropy shows that variationally estimating the entropy makes optimization more challenging. The lower average cost and large standard deviation hints at the presence of local minima in the cost landscape. The optimisation was performed using the LM-BFGS optimizer the results for different optimizers are qualitatively the same \cite{virtanen_scipy_2020}.}
    \label{fig:trans_invar}
\end{figure*}

In Fig.~\ref{fig:trans_invar} the final fidelity after optimization for six qubits is plotted at different inverse temperatures $\beta = 1/T$. We ran the optimization for ten initializations of the unitary parameters chosen from a Gaussian centered at 0. Panel (a) shows the Fidelity with the target state of the initialization that converged to the lowest cost. In the best case, all entropy estimation methods have a similar performance. Panel (b) shows the average fidelity and its standard deviation. The exact entropy and scaled subsystem entropy have similar performance on average as in the best case. The variational estimation performs significantly worse on average. 
This can be attributed to a cost landscape with many local minima. 

We conclude that for translationally invariant systems, where it is possible to use the scaled subsystem entropy, one should do so. It has two advantages over the variational entropy estimation. First, it is simpler to calculate as it only uses a single optimization loop and uses a circuit of lower depth. Second, its cost landscape is favorable compared to the variational entropy estimation. 
\section{Weak and strong symmetry conditions}\label{app:Symmetries}
\subsection{Preliminaries}
We first review relevant concepts from group and representation theory. These definitions can be found in any group theory textbook \cite{fulton_Representation_2004}. For a review in the context of quantum machine learning consider \cite{ragone_representation_2022}.

Symmetries are organized in groups. Recall the definition of a (finite) group. 
\begin{definition}[Group]
A group consists of a set $G$ equipped with a binary operation $\cdot : G \times G \rightarrow G$ that satisfies the following axioms:
\begin{enumerate}
    \item Composition is associative. For all $g,h,c \in G, \; (g\cdot h ) \cdot c = g \cdot (h\cdot c)$ 
    \item There exists an identity element $e$ in $G$ such that for any $g$ in $G$ $e, \cdot g = g \cdot e = g$.
    \item For every element $g$ in $G$ there exists an element $g^{-1}$ such that $g \cdot g^{-1} = g^{-1} \cdot g = e $.
\end{enumerate}
\end{definition}
When we consider how a symmetry group acts on a physical system, we require a mechanism to translate the abstract group into the physical Hilbert space that describes the system. This mechanism is known as the \textit{Representation}.
\begin{definition}[Representation]
A representation of a finite group G in a finite-dimensional complex vector space $V$ is a homomorphism from $G$ to the set of automorphisms on $V$
$
    R:G \rightarrow \text{Aut}(V) 
$
that respects the structure of the group. Specifically:
$
    R(a\cdot b ) = R (a) \circ R(b)
$
One refers to the dimension of the vector space $V$ as the dimension of the representation.
\end{definition}
It immediately follows that the representation of the identity element of the group is the identity 
\begin{equation}
    R(e) = \mathcal{I}.
\end{equation}
and 
\begin{equation}
    R(a^{-1}) = (R(a))^{-1}.
\end{equation}

Provided that $V$ is finite-dimensional one can introduce a basis and the representations are $n \times n $ invertible matrices. 
Two representations are said to be equivalent if they are related by a similarity transformation.

\begin{definition}[Subrepresentation]
    A subrepresentation of a representation $R$ on vector space $V$ is a subspace $W$ of $V$ such that for all $w \in W $ and $g \in G$ $ R(g)w \in W$.
\end{definition}

A representation is referred to as \textit{irreducible} if it does not have any nontrivial subrepresentations i.e. apart from itself and $\{0\}$.

\begin{definition}[Completely reducible]
    A representation $(R, V)$ is called completely reducible if invariant subspaces $V_1,\dots , V_n$ exist such that $V = V_1 \oplus \dots \oplus Vn$ and the subrepresentations $(R|_{V_i} , V_i)$ are irreducible.

\end{definition}
This decomposition of V is called a decomposition into irreducible representations or short irreps. We will denote the irreps as $\gamma_0,\gamma_1,\dots,\gamma_n$. 

\begin{definition}[Character]
    Let $V$ be a finite-dimensional vector space over $C$ and $R: G \rightarrow V $ a representation of group $R$ on $V$. Then the character $\chi_R (g)$  is a one dimensional representation $\chi_R : G \rightarrow C$ defined by $\chi_R(g) = \Tr [R_g ] $
\end{definition}

\begin{lemma}[Orthogonality of characters]\label{lemma:Orthogonality}
     Let $R$ an $R'$ be two distinct irreps of a finite-dimensional group with their respective characters $\chi_R$ and $\chi_{R'}$ then
\begin{equation}
\frac{1}{\vert G \vert } \sum_{g \in G} \chi_R(g) \overline{\chi_{R'}(g)} = \begin{cases} 1, & \text{if } R \text{ and } R' \text{ are equivalent}, \\ 0, & \text{otherwise}. \end{cases}
\end{equation}
Here, $\vert G \vert $ is the order of the group, and the sum is taken over all elements $g \in G$.
\end{lemma}

\subsection{Weak and strong symmetry conditions}
Following Ref. \cite{de_groot_symmetry_2022}, we introduce symmetry conditions for quantum channels. Consider a model described by Hamiltonian $H$ with symmetry group $G$ with \textit{unitary} representation $R_g$ for $g$ in $G$. 
\begin{definition}[Weak symmetry]
    A channel $\mathcal{E}$ satisfies the \textit{weak symmetry condition} if it commutes with the channel that represents the symmetry. The symmetry operation is represented by the map $\mathcal{R}_g(\rho) = R_g \rho R_g^\dag$, where $R_g$ is the unitary representation of $g \in G$.
    Specifically, for all $g$ in $G$
    \begin{equation}
        \mathcal{R}_g \circ \mathcal{E} \circ \mathcal{R}^\dag_g = \mathcal{E} 
    \end{equation}
\end{definition}

If we express the channel $\mathcal{E}$ in the Kraus representation, $\mathcal{E}(\rho)=\sum K_i \rho K_i^\dag$, the weak symmetry condition reads
\begin{equation}
    \sum (R_g K_i R_g^\dag) \rho ( R_g K_i R_g^\dag)^\dag = \sum K_i \rho K_i^\dag, \quad \forall g \in G.
\end{equation}

We conclude that if $\mathcal{E}$ meets the weak symmetry condition then the transformed operators $R_g K_i R_g^\dag$ are Kraus representations of the same channel. Therefore, they are unitarily equivalent (see Ref. \cite{nielsen_quantum_2010}) with unitary matrix $U^g$, such that
\begin{equation}
    R_g K_i R_g^\dag{} = \sum_j U_{ji}^g K_j ,\quad \forall i,g
\end{equation}

The unitary $U^g$ forms a representation of $G$. We show the composition property. Let $h$ be in $G$
\begin{align}
\sum_j U_{ji}^{gh} K_j = R_{gh} K_i R_{gh}^\dag{}= \\
R_{g} R_{h} K_i R_{h}^\dag{}R_{g}^\dag{} = \sum_{j,l} U_{li}^{h}  U_{jl}^{g} K_j .
\end{align}

We can choose a further unitary transformation that diagonalizes $U^g$. In this basis we represent the Kraus operators as $K_i^g$. 
\begin{equation}
    R_g K^g_i R_g^\dag = e^{i \theta_i(g)} K_i^g, \quad \forall i,g
\end{equation}
In this basis the action of $U^g$ is captured by the phases $\{\theta_i(g)\}$. The existence of this basis for all $g$ is equivalent to the weak symmetry condition.

If the phases $\theta_i(g)$ in the diagonal basis are independent of the Kraus operator $K_i$ we arrive at the strong symmetry condition.
\begin{definition}[Strong Symmetry]
    A Kraus channel with Kraus operators $\{K_i\}$ meets the strong symmetry condition if \begin{equation}
        R_g K_i R_g^\dag = e^{i\theta(g)} K_i, \forall i,g
    \end{equation}
\end{definition}
This definition is basis independent. To see this, suppose $\{K_i\}$ satisfies the strong symmetry condition. If we define a new set of transformed Kraus operators $K_j' = \sum_i \mu_{ji} K_i$, then under the symmetry operation, we have:
\begin{equation}
    R_g K_j' R_g^\dag = \sum_i \mu_{ji}  R_gK_i R_g^\dag = e^{i\theta(g)} K_k'.
\end{equation}
For unitary channels, the weak and strong symmetry conditions are equivalent because unitary channels have only one Kraus operator, and thus all Kraus operators trivially share the same phase.

\subsection{Symmetries of Lindbladian}\label{app:LindbladSym}
We consider quantum systems described by Lindblad master equations for the implementation of quantum channels. The dynamics of a quantum system coupled to a Markovian reservoir can be captured by a Master equation in Lindblad form
\cite{lindblad_generators_1976}:
\begin{equation}
    \mathcal{L}(\rho) = -\frac{i}{\hbar} \left[ H ,\rho \right] + \sum_i \left( L_i \rho L_i^\dag -\frac{1}{2}(L_i^\dag L_i \rho + \rho L_i^\dag L_i)\right).
\end{equation}
The Hamiltonian $H$ captures the coherent evolution, while the Lindblad jump operators $L_i$ capture the interaction with the reservoir.

\begin{definition}[Weak Lindbladian symmetry]
If a channel $\mathcal{E}$ generated by Lindbladian $\mathcal{L}$ satisfies a weak symmetry the jump operators satisfy
\begin{equation}
    \mathcal{R}_g \circ \mathcal{L} \circ \mathcal{R}^\dag_g = \mathcal{L} \quad \forall g \in G
\end{equation}
\end{definition}
This definition aligns with the weak symmetry condition for quantum channels because the evolution of the channel generated by the Lindbladian $\mathcal{E}_t = e^{t \mathcal{L}}$ will commute with $\mathcal{R}_g$ for all times $t$. 

A sufficient condition to satisfy a weak symmetry is that the Hamiltonian commutes with the symmetry (i.e., $\left[ H, R_g\right] = 0$) and the Lindblad jump operator $L_i$ anti commute with the symmetry (i.e., $\{ L_i,R_g \}$), for all $i$ and $g \in G$.

\begin{definition}[Strong Lindbladian symmetry]
If a channel $\mathcal{E}$ generated by Lindbladian $\mathcal{L}$ satisfies a strong symmetry the jump operators satisfy
    \begin{equation}
        R_g L_i = L_i R_g, \quad R_g H = H R_g  , \quad \forall i, g 
    \end{equation}
\end{definition}

Furthermore, the Kraus operators of channels generated by Lindbladian evolution that satisfy the strong symmetry commute with the symmetry representation i.e. $\theta(g) = 0 $.

To better understand the strong symmetry condition, we consider the evolution $\mathcal{E}_t$ for a small time interval $\delta t$. For small times, the evolution map $\mathcal{E}_{\delta t}$ can be expanded as:
\begin{align}
    \mathcal{E}_{\delta t} (\rho) &= \rho + \delta t \mathcal{L}(\rho) \nonumber\\
    &=  \rho +  \delta t \left( H_{\text{nh}} \rho + \rho H_{\text{nh}}^\dag  \right) + \sum_{j=1}^l L_j \rho L_j^{\dag{}}.
\end{align}
where we introduced the non-Hermitian Hamiltonian $H_{\text{nh}} = -i H - \frac{1}{2}\sum_{j = 1}^l L_j^\dag L_j$.
This channel has Kraus form with Kraus operators
\begin{align}
    K_0(\delta t ) &= \mathrm{I} + H_{\text{nh}}, \\
    K_{j\geq 0}(\delta t ) &= \sqrt{\delta t} L_j.
\end{align}
These operators clearly satisfy the strong symmetry condition for Kraus operators with $\theta(g) = 0 $. 

In summary, weak Lindbladian symmetry requires that the overall evolution commutes with the symmetry operations, allowing for some degree of flexibility in how the jump operators transform. In contrast, strong symmetry imposes stricter requirements, forcing both the Hamiltonian and the jump operators to commute with the symmetry, leaving no room for nontrivial phase factors.

\subsection{Symmetry sectors}\label{sec:symsec}
One can use symmetries to partition a quantum system's Hilbert space into distinct subspaces. The main advantage is that they allow us to reduce the dimension of the problem we have to consider greatly simplifying it \cite{marvian_symmetry-protected_2017,de_groot_inaccessible_2020}

The projector onto the symmetry sector of irrep $\alpha$
\begin{equation}
    \Pi_\alpha = \frac{1}{|G|} \sum_g \chi_\alpha (g) R_g.
\end{equation}

Here $\chi_\alpha(g)=\Tr\left[\gamma_\alpha(g)\right]$ is the character of irrep $\gamma_\alpha$. 
By construction, the projectors $\Pi_\alpha $ are mutually orthogonal and commute with the Hamiltonian. The image of each projector is called a \textit{symmetry sector} \cite{misguich_detecting_2007}. 

This decomposition has many advantages. For example, when diagonalizing the Hamiltonian, one can now work within these smaller subspaces. 

The overlap of state $\rho$ with symmetry sector $\alpha$ is
\begin{equation}
    p_\alpha  = \Tr\left[\Pi_\alpha \rho\right].
\end{equation}
Indeed probabilities sum to one 
\begin{align}
    \sum_\alpha p_\alpha(\rho) &= \frac{1}{|G|}\sum_g \Tr\left[ R_g \rho\right] \sum_\alpha \chi_\alpha(g) \nonumber\\
    &= \frac{1}{|G|}\sum_g \Tr\left[ R_g \rho\right] \delta_{g,e} |G| =1 .
\end{align}
We insert the character of the trivial representation $\chi_e(g) = 1$ and use the orthogonality of the characters in the second step. 
A strongly symmetric channel leaves the set $\{ p_\alpha \}$ invariant. 
\begin{align}
    p_\alpha( \mathcal{E}(\rho))& =\frac{1}{|G|} \sum_g \chi_\alpha (g) \Tr\left[R_g \sum_i K_i \rho K_i^\dag\right] \nonumber\\
       & =\frac{1}{|G|} \sum_g \chi_\alpha (g) \Tr\left[\sum_i e^{i \theta(g)} K_i  R_g  \rho K_i^\dag\right] \nonumber\\
    & =  \frac{1}{|G|} \sum_ge^{i \theta(g)} \chi_\alpha (g) \Tr\left[ \sum_i K_i^\dag K_i R_g \rho \right] \nonumber\\
    & =  \frac{1}{|G|} \sum_g\chi_{\tilde{\alpha}} (g) \Tr\left[ R_g \rho \right] \nonumber\\
    & =  p_{\tilde{\alpha}}(\rho).
\end{align}
Here we denote $e^{i \theta(g) } \chi_\alpha(g) = \chi_{\tilde{\alpha}}(g)$. 
If $\theta(g) = 0 $, i.e. the Kraus operators commute with the symmetry, then $\Tilde{\alpha} = \alpha$.  

We give an example to illustrate this permutation. Consider a group $G$ with representation on three qubits, $R_g = \{ \mathbb{I}\mathbb{I}\mathbb{I},XXX,X\mathbb{I}\mathbb{I},\mathbb{I}XX\}$. The channel $\mathcal{E}$ we consider has Kraus operators proportional to $Z\mathbb{I}\mathbb{I}$ and $ZZZ$ is strongly symmetric. To calculate the phases $\theta(g)$ associated to the group elements $g$ in $G$ we consider the commutation relations. The phases corresponding to the group elements are:  
 \begin{equation}
     \theta(\{111,XXX,X11,1XX\}) = \{0,\pi,\pi,0\}. 
 \end{equation}
The character table of the group is: 
\begin{equation}
\begin{tabular}{lllll}
\multicolumn{1}{l|}{}       & 111 & XXX    & X11 &  1XX \\\cline{1-5}
\multicolumn{1}{l|}{$\chi_0$}       &   1  &\,\! 1      & \,\!  1  &  \:1\\
\multicolumn{1}{l|}{$\chi_1$}       &   1  &    -1      & -1       &\:1  \\
\multicolumn{1}{l|}{$\chi_2$}       &   1  &  \,\!  1   &  -1      &  -1\\
\multicolumn{1}{l|}{$\chi_3$}       &   1  &    -1      &\,\! 1    & -1
\end{tabular}
\end{equation}
Acting with the channel $\mathcal{E}$ adjusts the characters as $e^{i \theta(g) } \chi_\alpha(g) = \chi_{\tilde{\alpha}}(g)$,
\begin{equation}
\begin{tabular}{l|llll}

       & 111 & XXX    & X11 &  1XX \\\hline
$\chi_{\tilde{0}}$       &   1  &    -1      & -1       &\:1   \\
$\chi_{\tilde{1}}$       &   1  &\,\! 1      & \,\!  1  &  \:1 \\
$\chi_{\tilde{2}}$        &   1  &    -1      &\,\! 1    & -1 \\
$\chi_{\tilde{3}}$       &   1  & \,\!1      &  -1      &  -1 \\
\end{tabular}
\end{equation}

As a consequence, the action of the channel swaps the population between the symmetry sectors:
\begin{equation}
    p_0(\mathcal{E}(\rho)) = p_1(\rho), \quad  p_2(\mathcal{E}(\rho)) = p_3(\rho).
\end{equation}
The action of the strongly symmetric channel permutes the irrep probabilities and leaves the set invariant.

We can view Gibbs state preparation through the lens of symmetry sectors. Given initial state $\rho_0$ with irrep probabilities $\{p_\alpha^0\}$ we want to apply operations such that the final irreps match that of the Gibbs state, $\{p_\alpha^\text{fin} \}=\{p_\alpha^\text{Gibbs} \}$. To this end, it is crucial that the channels used break the strong symmetry. At the same time, the final state $\rho^\text{fin}$ should commute with the symmetry and therefore the channel should weakly respect the symmetry. 

\section{Connection to \textit{An efficient and exact noncommutative quantum Gibbs sampler}} \label{ap:Connection}
Recently, the authors of Ref. \cite{chen_efficient_2023}  introduced the first efficiently implementable Gibbs sampler that can be applied to non-commuting Hamiltonians. They construct a Lindbladian $\mathcal{L}$ whose fixed point is the Gibbs state by employing a continuously parameterized set of jump operators. The energy resolution (and thereby time) required to implement each jump operator scales logarithmically with both the precision and the mixing time. The operators they construct ensure that the target Gibbs state $\rho_G$ is the steady state of the algorithm; $ \mathcal{L}\left[\rho_G\right] = 0 $. 
In most quantum Gibbs sampling algorithms this condition only holds approximately which connects fundamentally with the uncertainty principle.

Nevertheless, a physical implementation of the algorithm is out of reach for current quantum computers for which ancilla qubits are scarce and non-local connectivity is a challenge.
Inspired by this algorithm we present a Gibbs preparation algorithm that is accessible to near-term quantum computers and can even utilize noise intrinsic to the system. We prove that for high temperatures the evolution we implement can approximate the evolution presented in Ref. \cite{chen_efficient_2023}. For low temperatures, our algorithm can prepare the ground state just as a variational quantum eigensolver would. 
Finally, numerical simulations confirm the effectiveness across a range of temperatures.

We implement approximations of the jump operators of Ref. \cite{chen_efficient_2023} and use them to generate an evolution $\Tilde{\mathcal{L}}$. Here we bound the difference of the diamond norm $\norm{\mathcal{L}-\Tilde{\mathcal{L}}}_\diamond$ of the ideal evolution $\mathcal{L}$ and the approximate evolution $\tilde{\mathcal{L}}$. We first bound the difference of the jump operators $L(\omega)$ and $\Tilde{L}(\omega)$ in spectral norm and then relate this to the difference in the dynamics.
We first introduce the norms we use. The \textit{1-norm} (or trace norm) of a matrix $ A $ is defined as $ \|A\|_1 = \operatorname{Tr} \sqrt{A^\dagger A} $, which corresponds to the sum of its singular values. The \textit{spectral norm} (or operator norm) is given by $ \|A\| = \sup_{\|x\|=1} \|Ax\| $, which is the largest singular value of $ A$. A particularly important norm in quantum information is the \textit{diamond norm}, which extends the operator norm to completely bounded maps. For a linear map $ \Phi $, the diamond norm is defined as $ \|\Phi\|_\diamond = \sup_{X \neq 0} \frac{\| (\Phi \otimes \mathbb{I})(X) \|_1}{\|X\|_1} $, where the optimization is over matrices $ X $ acting on an extended space. This norm captures the worst-case distinguishability of a quantum channel when acting on part of an entangled state.

The ideal evolution can be expressed as
\begin{align}
\mathcal{L}  \left[ \cdot \right] &\coloneqq
 - i \left[ B, \cdot \right] 
 + \sum_{a \in A }\int_{-\infty}^\infty \gamma(\omega) \times \\
 & \left( A^a(\omega) ( \cdot ) A^a(\omega) ^\dag - \frac{1}{2}\{ A^a(\omega)^\dag A^a(\omega),\cdot \} \right).
\end{align}
In the Lindbladian $\mathcal{L}$ the transition weight $\gamma(\omega) = \text{exp}\left( -\frac{(\beta \omega +1)^2}{2}\right)$ weights the quantum operator Fourier transform of the jump $A^a$. 
\begin{align}
    A^a (\omega) \coloneqq \frac{1}{\sqrt{2\pi} }\int_{-\infty}^\infty e^{iHt} A^a e^{-iHt} e^{-i \omega t } f(t),
\end{align}
with the Gaussian filter $f(t) = \frac{e^{-t^2/\beta^2}}{\sqrt{\beta \sqrt{\pi/2}}}$. The filter suppresses large times motivating an expansion:
\begin{align}
    \Tilde{A}^a (\omega) =  \frac{1}{\sqrt{2\pi} }\int_{-\infty}^\infty \left( \sum_{m=0}^\infty \frac{t^m}{m!}\left[ iH, A^a\right]_m  \right) e^{-i \omega t } f(t).
\end{align}
Here; $\left[ iH,A ^a \right]_m =\left[ iH,\left[ iH,A ^a \right]_{m-1}\right] $ and $\left[ iH,A ^a \right]_0 = A ^a $.
We approximate the jump using the first two terms of the expansion: 
\begin{align}\label{eq:approxjumpcont}
    \tilde{A}^a (\omega) \coloneqq \frac{1}{\sqrt{2\pi} }\int_{-\infty}^\infty \left( A^a + t \left[ iH, A^a\right] \right) e^{-i \omega t } f(t).
\end{align}
We can upper bound the difference in the spectral norm of the exact and approximate jump operators
\begin{align}
    \delta = &\norm{\frac{1}{\sqrt{2\pi} }\int_{-\infty}^\infty \left( \sum_{m=2}^\infty \frac{t^m}{m!}\left[ iH, A^a\right]_m \right) e^{-i \omega t } f(t) \text{d} t }\\
    \leq& \sum_{m=2}^\infty \underbrace{\left|\frac{1}{\sqrt{2\pi} }\int_{-\infty}^\infty \left(  \frac{t^m}{m!} \right) e^{-i \omega t } f(t) \text{d} t  \right|}_{\text{I}_m} \norm{\left[ iH, A^a\right]_m}. \nonumber
\end{align}
The integral $\text{I}_m$ evaluates to 
\begin{align}
I_{2n} 
    = \underbrace{\left( \sqrt[4]{\frac{2 \beta^2}{\pi}}\right)}_{C_1} \frac{(\beta/2)^n}{n!}\text{1F1}\left(\frac{1}{2} + n, \frac{1}{2}, -\frac{1}{4} \beta^2 \omega^2\right),
\end{align}
for $m=2n$. 1F1 is a confluent hypergeometric function. We use the property $\text{1F1}[a, b, z] = e^z \text{1F1}[b - a, b, -z]$ to simplify the expression: 
\begin{align}
    I_{2n}  = C_1 e^{-\frac{1}{4} \beta^2 \omega^2}\frac{(\beta/2)^n}{n!}\text{1F1}\left(- n, \frac{1}{2}, \frac{1}{4} \beta^2 \omega^2\right).
\end{align}
Expressing the hypergeometric function in terms of Laguerre polynomials 
\begin{align}
    I_{2n} = C_1 e^{-\frac{1}{4} \beta^2 \omega^2}\frac{(\beta/2)^n}{n!} \frac{\Gamma(n+1) \Gamma(1/2)}{\Gamma(1/2 + n) } L_n^{-1/2}\left(\frac{1}{4} \beta^2 \omega^2 \right) 
\end{align}
We upper bound the Laguerre polynomial using $|L_n^\alpha(x)| \leq 2^{-\alpha} q_n e^{x/2}$ for $\alpha \leq - \frac{1}{2},n \in N$ and $x\geq 0$. With $q_n= \frac{(2n)!}{2^{n+1/2} n!} \approx \frac{1}{\sqrt[4]{2 \pi n !}}$
\cite{lewandowski_upper_1998}. This allows us to upper bound the even terms by
\begin{align}
    I_{2n} \leq C_1 e^{-\frac{1}{8} \beta^2 \omega^2} \sqrt[4]{\pi}\left(\frac{\beta}{2}\right)^2 \frac{(\beta/2)^{2n-1}}{(n-1)!}.
\end{align}

For $m= 2n+1$ one can express the integral as
\begin{align}
    I_{2n+1} = 
    \underbrace{ \left( \frac{\beta^{5/2}}{\sqrt[4]{2^5 \pi}}\right)}_{C_2} \omega \frac{\beta^{2n}}{n!}\text{1F1}\left(\frac{3}{2} + n, \frac{3}{2}, -\frac{1}{4} \beta^2 \omega^2\right) \\
    = C_2 \omega e^{-\frac{1}{4}\beta^2 \omega^2 } \frac{\sqrt{\pi}}{2} \frac{\beta^{2n}}{\Gamma(3/2 + n)} L_n^{1/2}(\frac{1}{4}\beta^2\omega^2). \nonumber
\end{align}
The Laguerre polynomial can be bounded using $|L_n^\alpha (x)| \leq \frac{(\alpha+1)^n}{n!} e^{x/2}$ for $\alpha,x\geq 0$, here we use the raising factorial; $(x)^n= (x)(x+1)\ldots(x+n-1)$ \cite{lewandowski_upper_1998}.
The bound for the odd terms
\begin{align}
     I_{2n+1}  \leq C_2 \omega  e^{-\frac{1}{4}\omega^2 \beta ^2 } \frac{\sqrt{\pi}}{2} \frac{\beta^{2n}}{n!}.
\end{align}
Finally, we bound the spectral norm of the $m$-fold commutator
$\norm{\left[ iH, A^a\right]_m} \leq 2^m\norm{H}^m \norm{A}$ using submultiplicativity of the spectral norm.
Combining the terms we find
\begin{align}
    \delta(\omega) & \leq (C_1 e^{-\frac{1}{8}\beta^2\omega^2} \sqrt[4]{\pi} (\beta\norm{H} )^2 \norm{A}  e^{(\beta\norm{H}) ^2} \nonumber\\
    &+ C_2 \omega e^{-\frac{1}{4}\beta^2\omega^2} \frac{\sqrt{\pi}}{2} \norm{|H|}_1 ^2 \norm{A}) e^{4 \beta^2\norm{H} ^2} , \nonumber\\
     &= \delta_e   e^{-\frac{1}{8}\beta^2\omega^2}  + \delta_o \omega e^{-\frac{1}{4}\beta^2\omega^2} .\label{eq:deltaomega}
\end{align}
We now relate the difference in spectral norm of the jump operators to the difference in diamond norm of the resulting Liouvillians.
We generalize Lemma 30 of Ref. \cite{rall_thermal_2023} to a continuous setting. 

\textit{Lemma:}
Assume $L(\omega)$ and $\Tilde{L}(\omega)$ are jump operators parameterized by $\omega$. They define Liouvillians $\mathcal{L}$ and $\mathcal{\Tilde{L}}$ via $\mathcal{L}(\cdot) = \int_{-\infty}^\infty \text{d}\omega \gamma(\omega) \mathcal{D}\left[ L(\omega)\right]$. Say the jumps satisfy
\begin{align}
    \norm{L(\omega)-\Tilde{L}(\omega)} \leq \delta(\omega),
\end{align}
and w.l.o.g. the norm $\norm{\Tilde{L}(\omega)}\leq 1$ is bounded by 1.
Then
\begin{align}
    \norm{\mathcal{L}-\Tilde{\mathcal{L}}}_\diamond \leq  \int_{-\infty}^\infty \text{d}\omega \gamma(\omega) \left( 4 \delta(\omega) + 2 \delta(\omega)^2\right).
\end{align}
\textit{Proof:}
Let $\rho$ be the matrix that achieves the induced trace norm of $\mathcal{L}-\Tilde{\mathcal{L}}$ i.e.
\begin{align}
    \norm{\mathcal{L}-\Tilde{\mathcal{L}}}_1 = \norm{\mathcal{L}(\rho)-\Tilde{\mathcal{L}}(\rho)}_1.
\end{align}
We denote $L_\omega \coloneqq L(\omega)$.
\begin{align}
    \norm{\mathcal{L}-\Tilde{\mathcal{L}}}_1 &\leq \int_{-\infty}^\infty \text{d}\omega \gamma_\omega \norm{L_\omega \rho  L^\dag_\omega -\Tilde{L}_\omega \rho  \Tilde{L}^\dag_\omega}_1  \nonumber\\
    &+ \frac{1}{2}\norm{ \Tilde{L}^\dag_\omega\Tilde{L}_\omega \rho -L^\dag_\omega L_\omega \rho   }_1 \nonumber\\
    &+\frac{1}{2}\norm{ \rho \Tilde{L}^\dag_\omega\Tilde{L}_\omega  -\rho L^\dag_\omega L_\omega   }_1 \nonumber\\
    &\leq \int_{-\infty}^\infty \text{d}\omega \gamma_\omega \norm{L_\omega} \delta_\omega + \norm{\Tilde{L}^\dag_\omega} \delta_\omega \nonumber\\
    &+ \norm{\Tilde{L}_\omega^\dag \Tilde{L}_\omega - L^\dag _\omega L_\omega } \nonumber\\
    &\leq \int_{-\infty}^\infty \text{d}\omega \gamma_\omega 2 \left( \norm{L_\omega} \delta_\omega + \norm{\Tilde{L}_\omega}\delta_\omega\right) \nonumber\\
    &\leq  \int_{-\infty}^\infty \text{d}\omega \gamma_\omega \left( 4 \delta_\omega + 2 \delta_\omega^2\right). \label{eq:Lemma30}
\end{align}
Here we use the fact $\norm{ABC }_1 \leq \norm{A}\norm{B}_1\norm{C}$.
The result can be extended to the diamond norm analogously to Ref. \cite{rall_thermal_2023}. This concludes the proof. 

We combine equation \eqref{eq:deltaomega} and the generalization of Eq. \eqref{eq:Lemma30} (Lemma 30 in Ref. \cite{rall_thermal_2023})
\begin{align}
    \norm{\mathcal{L}-\Tilde{\mathcal{L}}}_\diamond &\leq  \int_{-\infty}^\infty \text{d}\omega \gamma(\omega) \left( 4 \delta(\omega) + 2 \delta(\omega)^2\right) , \nonumber\\
    & \underset{\beta \rightarrow 0}{\approx} 4 \frac{ 2^{3/4} \norm{A}\norm{H}}{e^{1/12}} \sqrt{\frac{\pi}{3}} \beta^{\frac{3}{2}} + O(\beta^\frac{5}{2}) .\quad 
\end{align}
For small $\beta$ i.e. high temperatures the approximate evolution approaches the evolution as presented in Ref. \cite{chen_efficient_2023}
\begin{align}
    \norm{\mathcal{L}-\Tilde{\mathcal{L}}}_\diamond \approx \mathcal{O}(\beta).
\end{align}
The Gibbs state is an exact steady state of the ideal Liouvillian $\mathcal{L}$. The approximate evolution approaches the ideal evolution in $\beta$, therefore the Gibbs state is an approximate steady state of the approximate evolution for high temperatures. 
For low temperatures, the approximate evolution can deviate from the ideal evolution. 

Here we assumed that we could implement the continuous version of the trotterized jump (Eq. \eqref{eq:approxjumpcont}). In practice, we can only implement a discretization of it. The error that occurs due to discretization is bounded in Ref. \cite{chen_quantum_2023}, and can be used in conjunction with the triangle inequality to bound the error of the error of the trotterized and discretized evolution.

\section{Engineering of nonunitary multi-qubit operations}\label{ap:jump_experiment}
Nonunitary operations can be implemented in various ways. One common approach is to apply a unitary interaction to an enlarged system, followed by tracing out the ancillary degrees of freedom. However, this method introduces additional complexity by requiring additional resources.
Measurements also constitute a nonunitary operation and can be used to implement nonunitary gate operations \cite{terashima_nonunitary_2005}.

Here, we focus on nonunitary operations that can be implemented by means of dissipation engineering in systems described by a Lindblad master equation
\begin{align}\label{eq:lind}
    \Dot{\rho} &= - i [H,\rho] + \mathcal{D}_{\{L_j\}}(\rho)\\
    &=  - i [H,\rho] +  \sum_k L_k \rho L^\dag_k - \frac{1}{2}\left(L^\dag_k L_k\rho + \rho L^\dag_k L_k \right). \nonumber
\end{align}
Recently, it was demonstrated that nonunitary multi-qubit quantum gates can be realized by using dissipation \cite{zapusek_nonunitary_2023,van_mourik_experimental_2024}. Initially proposed for quantum dots interacting within a cavity \cite{zapusek_nonunitary_2023}, this scheme has been experimentally demonstrated using trapped ions \cite{van_mourik_experimental_2024}. Due to the general nature of the interactions involved, the approach is adaptable to other physical platforms, such as superconducting qubits \cite{reiter_steady-state_2013}. In this appendix, we describe how the jump operator for the TFIM and then the Heisenberg model can be implemented.
\subsection{Engineering of the single-qubit TFIM jump}\label{ap:jump_TFIM_experiment}
The jump operator used for the TFIM has a form 
\begin{align}
    L = \sqrt{p} \left(Z + d Y \right),
\end{align}
with $p$ and $d$ real parameters.
We can simplify the implementation of the jump operator by applying a unitary transformation. Application of a unitary operator $U$ to the input state of the dissipative evolution transforms the evolution to a new jump operator $\Tilde{L}$. To see this consider the dissipator: 
\begin{align}
    \mathcal{D}_L ( U \rho U^\dag{}) &= L U \rho U^\dag{} L^\dag{}  - \frac{1}{2} \{ L^\dag{}  L , U \rho U^\dag{} \} \nonumber\\ 
    &= U \Tilde{L} \rho \Tilde{L}^\dag U^\dag - \frac{1}{2} U  \{ \Tilde{L}^\dag{}  \Tilde{L} , U \rho U^\dag{} \} U^\dag  \nonumber\\
    &=  U \mathcal{D}_{\Tilde{L}} (\rho)  U^\dag{},
\end{align}
where $\Tilde{L} = U^\dag L U $. 
Concretely we apply a Hadamard and transform the jump to:
\begin{align}
    \Tilde{L} = HLH = \sqrt{p} \left(X - d Y \right).
\end{align}
We will engineer this coupling in a system that consist of four levels $ \{ \ket{0},\ket{1},\ket{e},\ket{f} \}$ coupled to an harmonic oscillator degree of freedom with ladder operators $a$. We assume that the system dynamics can be described by a master equation in Lindblad form. 

The system Hamiltonian contains sideband interactions $H_s$ coupling levels $\ket{e}$ and $\ket{f}$ to the oscillator and a weak drive $V$ referred to as the probe. 
\begin{align}
  H = H_s + V,
\end{align}
where
\begin{align}
    H_s = g ( a^\dag{} \ketbra{1}{e}+ a^\dag{} \ketbra{0}{f} + h.c ), 
\end{align}
and 
\begin{align}
    V = \Omega (e^{-i \phi} \ketbra{e}{0} + e^{i \phi} \ketbra{f}{1} + h.c).
\end{align}
In addition, we consider cooling of the harmonic oscillator 
\begin{align}
    L = \sqrt{\kappa} a.
\end{align}
In the limit that the cooling is stronger than the excitation, $\kappa \gg \Omega$, one can describe the system dynamics with effective jump operators on the $\{\ket{0},\ket{1}\}$ manifold \cite{reiter_effective_2012}. The jump takes the form: 
\begin{align}
    L_\text{eff} = \frac{\sqrt{\kappa}}{2g}(\Omega e^{i \theta} \ketbra{0}{1} + \Omega e^{i \phi} \ketbra{1}{0}).
\end{align}
Choosing 
\begin{align}
 \phi = \arctan(d) ,\quad \frac{\sqrt{\kappa}\Omega}{2 g } = \sqrt{p(1+d^2)},
\end{align}
yields an effective jump operator of the form $\Tilde{L}$.
\subsection{Engineering of the two-qubit Heisenberg jumps}\label{ap:jump_Heis}
\begin{figure}
    \centering
    \includegraphics[width=\linewidth]{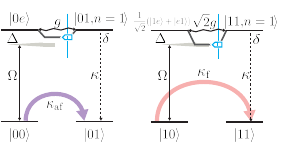}
    \caption{Engineering of nonunitary multi-qubit operation. Level scheme of the effective two-body jump operators. State $\ket{0}_2$ is off-resonantly excited to $\ket{e_2}$ . The excited state is shifted conditioned on the state of the first qubit due to the strong coupling $g$. Adjusting the detunings $\delta$ and $\Delta$ the relative excitation strength of states $\ket{00}$ and $\ket{10}$ can be controlled. Subsequently, the operation is completed by the decay of the oscillator.}
    \label{fig:Heist_jumps_app}
\end{figure}

The jump operators we use to prepare thermal states of the Heisenberg model are: 
\begin{align}
    L_{\kappa,0} ^\text{eff} &= \sqrt{\kappa_\text{f}}\ketbra{11}{10} + \sqrt{\kappa_\text{af}}\ketbra{01}{00},\\
    L_{\kappa,1}^\text{eff} &= \sqrt{\kappa_\text{f}}\ketbra{00}{01} + \sqrt{\kappa_\text{af}}\ketbra{10}{11}.
\end{align}
We consider a physical system of two qubits and four levels $\{ \ket{0}, \ket{1}, \ket{e},\ket{f}\}$ coupled to an oscillator mode with ladder operator $a$ and $b$. We will discuss the implementation of $L_{\kappa,0} ^\text{eff}$, the jump $ L_{\kappa,1}^\text{eff}$ can be implemented analogously.
The system Hamiltonian contains sideband interactions $H_s$ coupling levels $\ket{e}$ to the oscillator and a weak drive $V$ referred to as the probe. 
\begin{align}
  H =  H_s + V  + \delta a^\dag a +  \Delta  \sum_j\ketbraind{e}{j}{e} ,
\end{align}
where
\begin{align}
    H_s = g \sum_{j \in \{1,2\}} \left( a^\dag \ketbraind{1}{j}{e} + a \ketbraind{e}{j}{1}\right),
\end{align}
and
\begin{align}
    V = \frac{\Omega}{2} \left( \ketbraind{e}{2}{0} + \ketbraind{0}{2}{e} \right).
\end{align}
The dissipative contribution to the operation is cooling of the oscillator mode:
\begin{align}
    L = \sqrt{\kappa} a.
\end{align}
We show the action of the couplings on two-qubit states in Fig. \ref{fig:Heist_jumps_app}.
In the limit that the cooling is stronger than the excitation, $\kappa \gg \Omega$, one can describe the system dynamics with effective jump operators on the $\{\ket{0},\ket{1}\}$ manifold \cite{reiter_effective_2012}. The jump takes the form: 

\begin{align}
L_{\kappa,0} ^\text{eff} &= \sqrt{\kappa_\text{f}}\ketbra{11}{10} + \sqrt{\kappa_\text{af}}\ketbra{01}{00},
\end{align}
with
\begin{align}
    \sqrt{\kappa_\text{f}}&= \sqrt{\kappa}\frac{\Omega}{2}\frac{\tilde{\delta}}{\tilde{\delta} \Delta - g^2 },  \\
    \sqrt{\kappa_\text{af}}&= \sqrt{\kappa}\frac{\Omega}{2}\frac{\tilde{\delta} \Delta - g^2}{\tilde{\delta} \Delta^2 - 2g^2 \Delta }.
\end{align}
Tuning the product of the detunings $\delta \Delta$ relative to $g$ allows to tune the relative strength of the decay processes. To adjust how much population decays the evolution time or drive strength $\Omega$ can be adjusted.
In addition there an effective Hamiltonian acts on the ground state manifold 
\begin{align}
    H_\text{eff} =& \left(\frac{\Omega}{2}\right)^2 2 \Re{\frac{\tilde{\delta}}{\tilde{\delta} \Delta - g^2 } }\ketbra{00}{00} \\
    & + \left(\frac{\Omega}{2}\right)^2 2 \Re{ \frac{\tilde{\delta} \Delta - g^2}{\tilde{\delta} \Delta^2 - 2g^2 \Delta }} \ketbra{10}{10}.
\end{align}
The effect of the Hamiltonian is undesired but can be echoed away by flipping the phase of $\delta$ and $g$ after half the evolution. 

\bibliographystyle{apsrev4-2}
\bibliography{references_VQT_abbrev}

\end{document}